\documentclass[aps,prc,reprint,superscriptaddress,nofootinbib]{revtex4-2}

\usepackage{graphicx} % Include figure files
\usepackage{xcolor}
\usepackage{ulem}
\usepackage{hyperref} % add hypertext capabilities
\hypersetup{colorlinks=true, citecolor=blue, urlcolor=blue, linkcolor=blue}

\begin{document}
\newcommand{\elem}[3]{^{#1}_{#2}\mathrm{#3}}
\newcommand{\units}[1]{\ \mathrm{[#1]}}
\newcommand{\unitsn}[1]{\ \mathrm{#1}}
\newcommand{\bv}[1]{\mathbf{#1}}

%\preprint{}

\title{Smoothing of one- and two-dimensional discontinuities in potential energy surfaces}


\author{N.-W. T. Lau}
\email[]{ngee-wein.lau@anu.edu.au}
\affiliation{Department of Fundamental and Theoretical Physics, Research School of Physics, Australian National University, Canberra, Australian Capital Territory 2601, Australia}
\affiliation{Department of Nuclear Physics and Accelerator Applications, Research School of Physics, Australian National University, Canberra, Australian Capital Territory 2601, Australia}

\author{R. N. Bernard}
\affiliation{Department of Fundamental and Theoretical Physics, Research School of Physics, Australian National University, Canberra, Australian Capital Territory 2601, Australia}

\author{C. Simenel}
\affiliation{Department of Fundamental and Theoretical Physics, Research School of Physics, Australian National University, Canberra, Australian Capital Territory 2601, Australia}
\affiliation{Department of Nuclear Physics and Accelerator Applications, Research School of Physics, Australian National University, Canberra, Australian Capital Territory 2601, Australia}


\date{\today}

\begin{abstract}
\begin{description}
\item[Background] The generation of potential energy surfaces is a critical step in theoretical models aiming to understand and predict nuclear fission. Discontinuities frequently arise in these surfaces in unconstrained collective coordinates, leading to missing or incorrect results.
\item[Purpose] This work aims to produce efficient and physically-motivated computational algorithms to refine potential energy surfaces by removing discontinuities.
\item[Method] Procedures based on tree-search algorithms are developed which are capable of smoothing discontinuities in one and two-dimensional potential energy surfaces while minimising their overall energy.
\item[Results] Each of the new methods is applied to smooth candidate discontinuities in $\elem{252}{}{Cf}$, $\elem{222}{}{Th}$ and $\elem{218}{}{Ra}$. The effectiveness of each case is analysed both qualitatively and quantitatively. The one-dimensional method is also compared to the adiabatic and linear interpolation approaches which are commonly used to remove discontinuities.
\item[Conclusions] The smoothing methods presented in this work are resource-efficient and successful for one- and two-dimensional discontinuities; they will improve the fidelity of potential energy surfaces as well as their subsequent uses in beyond mean-field applications. Complex discontinuities occurring in higher dimensions may require alternative approaches which better utilise prior knowledge of the potential energy surface to narrow their searches.
\end{description}
\end{abstract}

%\keywords{}

\maketitle

\section{\label{sec:intro}Introduction}

Although our understanding of nuclear fission has grown significantly since its discovery in the 1930s, it remains a highly active field of research today. As experimental measurements of fission are conducted with increasing precision, our theoretical models of the structure and dynamics of the nucleus must be further developed to verify and interpret these new observations. \\

A standard approach to modelling nuclear fission involves two distinct stages of calculations. First, the static relations between the nuclear shape and its binding energy must be obtained in the form of a potential energy surface (PES). Next, the dynamic evolution of the system from the compact nucleus to scission must be simulated and analysed. See Refs. \cite{schunck2016, bender2020, bulgac2020} for recent reviews of the methods for each stage and their applications. This work is focused on the generation of PESs using self-consistent Hartree-Fock-Bogoliubov (HFB) theory, a mean-field approach which calculates the ground-state energy and configuration of a nucleus from an effective nuclear interaction and shape constraints. \\

In addition to providing a qualitative description of fission modes, PESs are an essential component in the Time-Dependent Generator Coordinate Method (TDGCM) \cite{wawong1975, reinhard1983, verriere2020}, which is used to explore fission dynamics by simulating the time evolution of the nuclear wavefunction. This beyond mean-field method produces a probability distribution of scission configurations for a fissioning nucleus, allowing the calculation of the charge-mass distributions and total kinetic energy of the products. Initial applications of the TDGCM relied on the Gaussian overlap approximation (GOA) to simplify the onerous task of calculating integral kernels between states \cite{berger1984, berger1989, goutte2005}, and its use has continued since \cite{regnier2016, regnier2019}. However, this approximation is only valid when the PES is smooth and well-behaved, which is often not the case. In particular, it is common for a PES to contain discontinuities, which are regions where one or more of the nucleus' collective degrees of freedom changes by an unphysically large amount, sometimes accompanied by a sudden shift in energy \cite{dubray2012}. Applying the GOA to such a PES is poorly justified; however, the discontinuities themselves also present a barrier to time evolution in the exact TDGCM framework. Therefore a systematic procedure is needed to smooth out discontinuities and restore the physicality of the affected regions in the PES. Such a method will not only allow applications of the GOA to be better motivated, but also facilitate future TDGCM calculations without the GOA. \\

The goal of this work is to present and evaluate concrete methods to smooth discontinuities effectively without excessive computational cost. Section \ref{sec:background} will briefly introduce the required aspects of PES generation using the self-consistent HFB model, followed by an explanation of the causes and consequences of discontinuities in PESs and some general approaches to resolving them. Section \ref{sec:method} presents two algorithms which are able to smooth discontinuities in one-dimensional (1D) and two-dimensional (2D) PESs. These methods are put to the test in Sec. \ref{sec:results} against discontinuities in the PESs of $\elem{252}{}{Cf}$, $\elem{222}{}{Th}$, and $\elem{218}{}{Ra}$. Section \ref{sec:conclusions} summarises our findings and suggests possible directions for future work.

\section{\label{sec:background}Background}
\subsection{Self-consistent Hartree-Fock-Bogoliubov model}

\subsubsection{\label{sec:background_theoretical}Theoretical framework}

The nuclear density can be effectively represented by an infinite expansion of multipole moments $Q_{lm}$. In the current work, only the axially symmetric multipole moments $Q_{l0}$ are considered, defined as
\begin{equation}
Q_{l0} \equiv \langle \hat{Q}_{l0} \rangle = \sqrt{\frac{2l + 1}{4\pi}} \int d\bv{r}\ \rho(\bv{r}) P_l(\cos\theta), \label{eq:multipole}
\end{equation}
where $P_l$ is an ordinary Legendre polynomial. Variations of the potential energy with non-axial multipole moments ($m \neq 0$) are generally minimal, especially for nuclei in the actinide and transactinide regions. The multipole moment $Q_{22}$ associated with triaxiality is an exception, as it is known to lower the height of the first fission barrier in some systems \cite{larsson1972,schunck2014}. Extensions of this work to permit variations in $Q_{22}$ in the appropriate regions should be straightforward. \\

With this in mind, the \textsc{HFBaxial} program \cite{HFBAxial} was chosen as an implementation of Hartree-Fock-Bogoliubov theory which assumes axial, time-reversal and simplex symmetries. The effective nuclear interaction used to derive a mean-field interaction is the phenomenological Gogny interaction \cite{gogny1975,decharge1980}; calculations in this work were performed using the D1S parametrisation \cite{berger1984,berger1989,berger1991,robledo2019}. \\

By solving the HFB equation for the Hamiltonian derived from the chosen interaction, a solution wavefunction can be found for the specified constraints (a complete description of the theory can be found in Chap. 7 of \cite{ringandschuck}). The use of this method is built on an adiabatic approximation: the collective motion of the nucleus towards scission occurs on a longer timescale than the local dynamics of the nucleons, so the nucleus is assumed to evolve through a series of local ground states for each set of collective coordinates \cite{bender2020}. \textsc{HFBaxial} uses the gradient method \cite{robledo2011} to iteratively reach a ground state solution expressed in a finite basis of deformed harmonic oscillator states. The basis is deformed from the spherical shape by specifying the length and number of shells $N_\perp$ and $N_z$ in the axial and $z$ directions respectively. The program may also optionally adjust the axis lengths $b_\perp$ and $b_z$ of the deformed harmonic oscillator to best represent each solution wavefunction. \\

Under the assumption that the nuclear shape varies smoothly across the PES, \textsc{HFBaxial} is used to calculate the surface point-by-point, taking the final solution for each point as the initial state for an adjacent one. The direction in which solutions are propagated in this way is chosen by the user. While the resulting PES should ideally be independent of the propagation direction, this is often not the case, and solutions which differ depending on the direction of propagation are strong qualitative indicators of a discontinuity.

\subsubsection{Potential Energy Surfaces}

With a coordinate system for the nuclear density and a framework to obtain the wavefunction and potential energy for a nuclear configuration expressed in these coordinates, a PES can be generated. The ideal PES of a nucleus is infinite-dimensional and describes its potential energy in any possible configuration, giving full information on the ground state, fission barriers, and scission configurations. In practice, PES calculations are limited to a finite number of coordinates, so only a subspace of the ideal PES may be explored. In nuclear fission, the most important multipole moments are considered to be $Q_{20}$, $Q_{30}$, and $Q_{40}$,\footnote{Close to the scission line, there is evidence \cite{schunck2014,younes2009} suggesting that an additional parameter $Q_N$, constraining the density of particles in the neck, must be considered independently of $Q_{40}$ to correctly describe the shape of the pre-fragments.} representing axial elongation, mass asymmetry and neck shape respectively. Effects of higher-order coordinates on the potential energy are assumed to be negligible. \\

There are a few typical choices for coordinates when constructing a PES, each of which provides different information on the fission dynamics of the nucleus:
\begin{itemize}
\item $Q_{20}$ only: This results in the ``1D fission path'' of the nucleus, showing the ground state and any barriers that the nucleus must overcome to proceed to fission. The dominant fission mode of a nucleus can be predicted based on whether the value of $Q_{30}$ remains at zero (symmetric) or increases with deformation (asymmetric).
\item $Q_{20}$ and $Q_{30}$: This generates a two-dimensional PES which describes symmetric and asymmetric fission valleys that the nucleus could take towards fission, as well as the scission line. From these coordinates, the charge and mass distributions as well as the total kinetic energies of scission configurations can be determined with a time-dependent method such as TDGCM.
\item $Q_{20}$ and $Q_{40}$: Symmetric fission modes of a nucleus may be investigated by constraining all odd multipole moments to zero and generating a 2D PES in these two coordinates.
\item $Q_{20}$ and $Q_{22}$: Triaxial deformations corresponding to the $Q_{22}$ multipole moment are known to lower the heights of first fission barriers in actinide nuclei \cite{larsson1972,schunck2014}. When precise barrier heights are needed, a small PES can be calculated around the barrier with a triaxially-symmetric basis to correct a larger axially-symmetric 2D PES.
\end{itemize}

When calculating a PES, the constrained coordinates $Q_{lm}$ must be discretised onto a mesh. The mesh sizes for all calculations in this work were $2 \unitsn{b}^{l/2}$ in $Q_{20}$, $Q_{30}$ and $Q_{40}$.

\subsection{Discontinuities in PESs}
\subsubsection{Causes and characteristics}

\begin{figure}
\includegraphics[width=8.6cm]{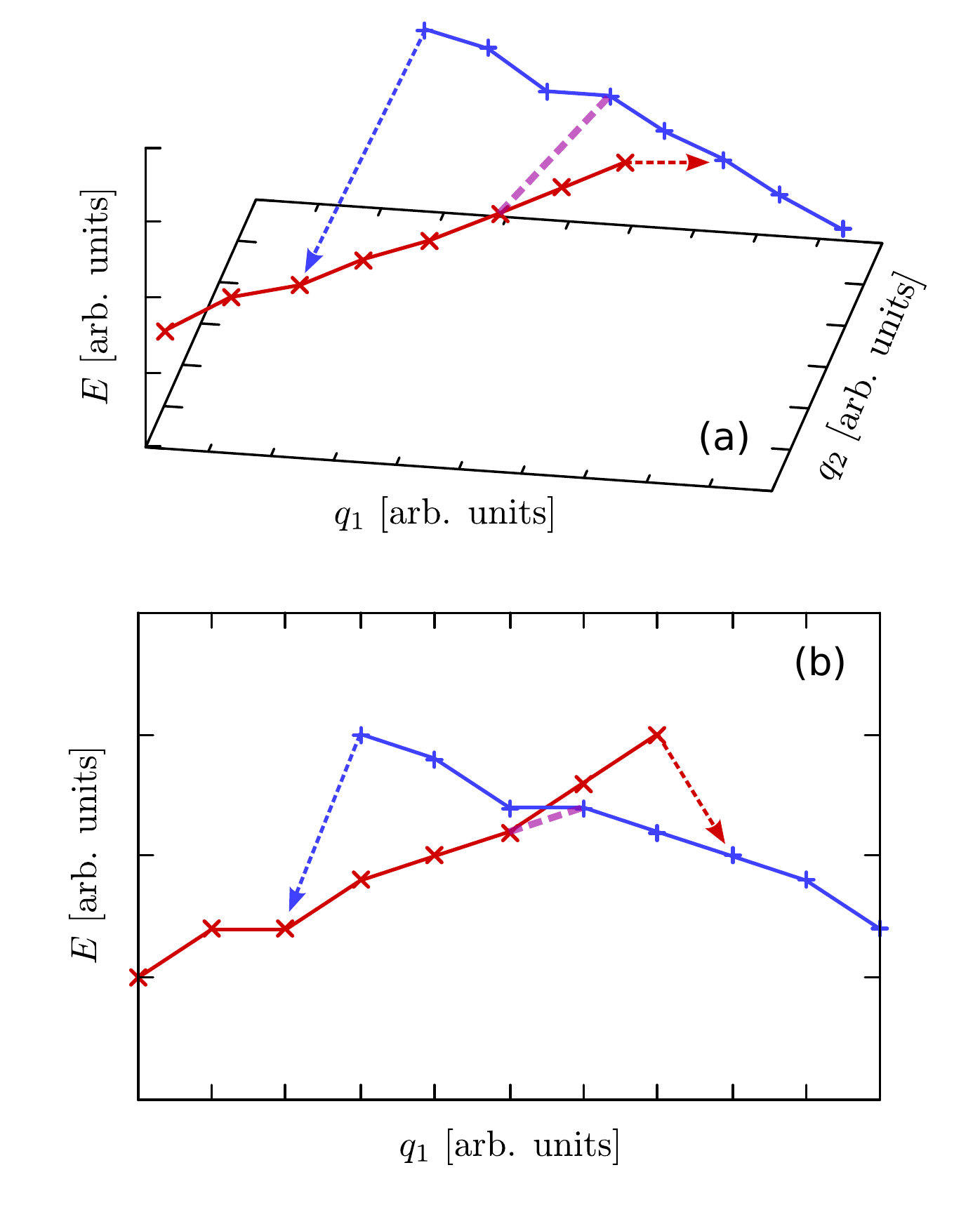}
\caption{\label{fig:discontinuity_projection}An example discontinuity between two local minima on a PES, represented by red and blue paths. (a) The paths are plotted in the arbitrary $q_1$-$q_2$ coordinate space, showing the large separation in $q_2$ between them. However, if the 1D fission path is calculated in $q_1$, the 1D projections of the paths intersect as shown in (b). HFB solutions propagating along the local minima will jump from one to the other as shown by the red and blue dashed arrows, while the adiabatic solution following the global minimum will cross between the paths at the purple dashed line. All three crossings exhibit large, discontinuous changes in the unconstrained $q_2$ coordinate, even though the adiabatic path appears to be smooth in energy.}
\end{figure}

The fundamental cause of discontinuities in any PES is the limited number of degrees of freedom (DOFs) which may be constrained in its generation. In an infinite-dimensional PES constraining all possible DOFs, which would take an infinite time to compute, any two points with similar energies lying in different fission valleys are distinguishable by a difference in one or more coordinates (which could be multipole moments $Q_{lm}$ or any other DOFs) as illustrated in Fig. \ref{fig:discontinuity_projection}(a). However, if a finite-dimensional PES is calculated without constraining the coordinates by which the points differ, they will be projected into the reduced space, shown by Fig. \ref{fig:discontinuity_projection}(b), such that they appear to be close to each other. The HFB algorithm may then choose to jump from one valley to the other between steps, producing a smooth change in the constrained DOFs but a discontinuous shift in one or more unconstrained coordinates. It should be noted that genuine discontinuities are not a consequence of the discretisation of the PES, and will not be eliminated by improving the resolution of the calculation. \\

As a variational method, self-consistent HFB searches for stationary points in energy by adjusting the nuclear degrees of freedom, subject to any imposed constraints. When propagating solutions in HFB calculations, this has the effect of favouring solutions that follow local energy minima, even if there are lower-energy minima elsewhere in the space. As a result, discontinuities in the nuclear density may also be accompanied by a sudden drop in potential energy when the algorithm jumps between different fission valleys. Propagating solutions in a different direction across such regions will frequently produce different results, as the method's tendency to follow local minima will result in a transition between the valleys at different locations. \\

The most obvious issue of a discontinuous PES is that it does not represent all of the configurations occupied by the fissioning system. The state of a nucleus must follow a smooth path across the PES, without sharp changes in any of its degrees of freedom. This gives rise to a greater concern: in general, for a nucleus to travel between two points on a PES it must cross the intervening energy surface. However, when crossing a discontinuity, only the start and end points are known, while the variation in energy between them due to the discontinuous coordinate is not defined by the PES. This has especially serious consequences when discontinuities occur close to a fission barrier, as the height of the barrier may be hidden by a discontinuity. \\

As understanding has grown of the impact that discontinuities can have on the reliability of PES data, recent efforts have been made to characterise them, both generally \cite{dubray2012} as well as in specific cases \cite{regnier2016,regnier2019,zdeb2021}. These studies suggest that discontinuities occur frequently on one- and two-dimensional PESs. The extent to which PESs are affected varies depending on the proximity of discontinuities to important surface features such as barriers or valleys, as well as whether the discontinuities cut across the expected fission paths or run parallel to them. \\

With the goal of modelling time evolution across the PES in mind, discontinuities cause an additional complication. The evolution between two states in the TDGCM framework depends on the Hamiltonian and overlap integral kernels.\footnote{For a formal explanation of TDGCM, see Refs. \cite{reinhard1983, verriere2020}. Reference \cite{wawong1975} and Chap.~10 of Ref. \cite{ringandschuck} describe the time-independent formulation.} The latter quantity, subsequently referred to as the ``overlap,'' is defined for two states as the inner product
\begin{equation}
\mathcal{N}(\bv{q}, \bv{q}') = \big\langle \varphi(\bv{q}) \big| \varphi(\bv{q}') \big\rangle, \label{eq:overlap}
\end{equation}
where $\varphi$ is the HFB wavefunction of the nucleus, and $\bv{q},\bv{q}'$ are the collective coordinates $(Q_{20}, Q_{30}, \dots)$ of the states. The wavefunctions of two states across a discontinuity represent significantly different nuclear configurations, which results in their overlap going to zero. This causes a singularity when attempting to solve the Hill-Wheeler equation in TDGCM, prohibiting the nucleus from evolving in time between the discontinuous states. Discontinuities situated near or across a fission valley therefore present a serious problem, as they may interfere with the evolution of the nuclear state towards a scission configuration.

\subsubsection{\label{sec:background_discontinuities_numerical}Numerical identification}

Although many discontinuities may be easily identified by looking at a plotted PES, it is useful to be able to locate discontinuities without relying on visual analysis. One quantity that can be used for this purpose is the overlap introduced in Eq. \ref{eq:overlap}. Since the inner product of two states is an indication of their similarity, it logically follows that a low or zero overlap between two adjacent points on a PES corresponds to a large change in nuclear configuration and hence a discontinuity \cite{verriere2017,verrierephd2017}. \\

Overlaps in this work have been computed using the Onishi formula \cite{onishi1966}. However, these calculations assume that the two states $|q\rangle$ and $|q'\rangle$ are expressed in the same basis. While the number of shells in the harmonic oscillator basis used by \textsc{HFBaxial} is fixed throughout the PES, the deformation of the basis parametrised by the oscillator lengths, as mentioned in Sec. \ref{sec:background_theoretical}, may vary between states. If the oscillator lengths between two states differ significantly, the overlaps calculated with the Onishi formula will erroneously go to zero. The Onishi formula could be extended by involving a unitary transformation between the different bases \cite{robledo1994}, but this was not deemed necessary in the current work. \\

The second quantity used to identify discontinuities is the density distance $D_{\rho\rho'}$, given in Refs. \cite{dubray2012,zdeb2021} as
\begin{equation}
D_{\rho\rho'} = \int \big|\rho(\bv{r}) - \rho'(\bv{r})\big| \,d\bv{r}. \label{eq:dens_dist}
\end{equation}
Instead of an inner product between states, this measures the overall difference between the densities $\rho$ and $\rho'$ of two adjacent nuclear configurations over the coordinate space. As such, the density distance is proportional to the change in nuclear configuration, and so a high density distance indicates the presence of a discontinuity. In Ref. \cite{dubray2012}, it is suggested on an empirical basis that a continuous PES will have a maximum density distance between closest neighbours of $D_\mathrm{max} < 2$. However, this limit is reliant on the choices of basis wavefunctions and coordinate mesh size, and it is additionally observed that variation of the oscillator lengths of the basis inflates the density distances. Because of these factors, an absolute upper threshold on the density distance for continuous neighbours was not imposed.

\subsubsection{\label{sec:smoothing_discontinuities}Smoothing discontinuities}

For discontinuities in 1D fission paths, there are two common simple approaches to resolve the problem on the surface level:
\begin{itemize}
\item In the adiabatic method, a new path is generated by selecting the lowest-energy value of the discontinuous coordinate from the 2D PES for each step of the independent coordinate across the discontinuity.
\item In the linear interpolation (``linear'') method, a path is generated over the offending region by linearly interpolating values of the discontinuous coordinate over the dependent coordinate.
\end{itemize}
While both methods can be used to obtain improved paths across a one-dimensional discontinuity, neither are fully satisfying. The adiabatic method usually produces paths continuous in energy, but they almost always contain discontinuities in the nuclear collective coordinates \cite{dubray2012}, meaning that the discontinuities are not truly resolved. While the linear interpolation method yields paths which are smooth in both energy and nuclear configuration, the endpoints for the interpolation must be chosen by the user. Furthermore, there is no effort to minimise the energy apart from the user's common sense in aligning the paths with fission valleys on the 2D PES, meaning that this approach does not uphold the HFB method's assumption of adiabaticity. \\

Two procedures to smooth discontinuities more comprehensively are suggested in Ref. \cite{dubray2012}. The first is to simply extend the full PES calculation into additional dimension(s) to include the coordinates in which its discontinuities occur. While this is guaranteed to remove the discontinuities, doing so greatly increases the resources required for the PES calculation, and may worsen the complexity of any applications that use the PES data. Higher-order multipole moments often have little impact on the PES outside of discontinuities, which makes such calculations potentially wasteful. \\

The second approach is called the ``connecting points'' method. Instead of extending the full PES into additional dimensions, a small higher-dimensional PES is calculated in the region of the discontinuity. Then a suitable pathfinding algorithm is applied to find the lowest-energy continuous path or surface across the region. \\

At the time of writing, these conceptual methods have not been widely applied to PES data. Zdeb and collaborators \cite{zdeb2021} recently published detailed analyses of the fission barriers, valleys, and discontinuities of $\elem{252}{}{Cf}$ and $\elem{258}{}{No}$ using three-dimensional $Q_{20}$-$Q_{30}$-$Q_{40}$ PESs. Their work demonstrates the effectiveness of using higher-dimensional calculations to describe and understand discontinuities. Additionally, Schunck \textit{et al.} \cite{schunck2014} and Younes and Gogny \cite{younes2009} have previously explored the possibility of smoothing two-dimensional PESs near the scission line by calculating a smaller PES to incorporate the $Q_N$ degree of freedom, which describes pre-scission configurations more accurately than the $Q_{40}$ multipole moment. The latter works bear some resemblance to the ``connecting points'' method, but they avoid higher-dimensional calculations by keeping some existing coordinates fixed to generate the PES in $Q_N$. This means that $Q_N$ does not vary alongside the original coordinates, and hence a truly continuous PES is not obtained. \\

The current work explores the possibility of using higher-dimensional calculations to smooth discontinuities in the original PESs using the ``connecting points'' method. A set of criteria was devised to evaluate the suitability of potential smoothing methods:
\begin{itemize}
\item The method should find the path or surface with the minimum total energy. It is acceptable for points to deviate from local minima of the fission surface, if doing so would lower the overall energy.
\item The path or surface generated by the method should be smooth in energy as well as in the coordinates $Q_{lm}$ either constrained by the PES or involved in the discontinuity.
\item The overlaps [Eq. (\ref{eq:overlap})] and density distances [Eq. (\ref{eq:dens_dist})] between adjacent points on the generated path or surface should show significant improvements when compared to the original PES. In particular, there should be no adjacent points with near-zero overlap between them.
\item The method should be as efficient as possible in its use of computing time and memory, to facilitate its general application to many discontinuities and PESs at a reasonable cost.
\end{itemize}
With these criteria in mind, the next section will introduce new methods to smooth discontinuities in one- and two-dimensional PESs. It should be noted that the problem of smoothing discontinuities does not generally admit a unique solution. If this is desired, further conditions such as minimising the barrier height over the smoothed region could be included.

\section{\label{sec:method}Methodology}
\subsection{Dynamic Programming Method for 1D path smoothing}

The Dynamic Programming Method (DPM) is a search algorithm developed in Ref. \cite{baran1981} and applied in Ref. \cite{sadhukhan2013} to determine the path of least action over a fission barrier in a PES. In the quasi-classical Wentzel-Kramers-Brillouin approximation, the spontaneous fission half-life can be estimated from the classical action of the optimal path. \\

The DPM algorithm is based on a conventional breadth-first tree search (BFS). The search occurs within a grid of points in two coordinates, defined as $x$ and $y$, which discretises the two-dimensional PES: each point $(x,y)$ has an associated energy, wavefunction, and collective inertia. When provided with initial and final coordinates $I \equiv (x_i, y_i)$ and $F \equiv (x_f, y_f)$, the method calculates paths across the search space between them. \\

\begin{figure}
\includegraphics[width=8.6cm]{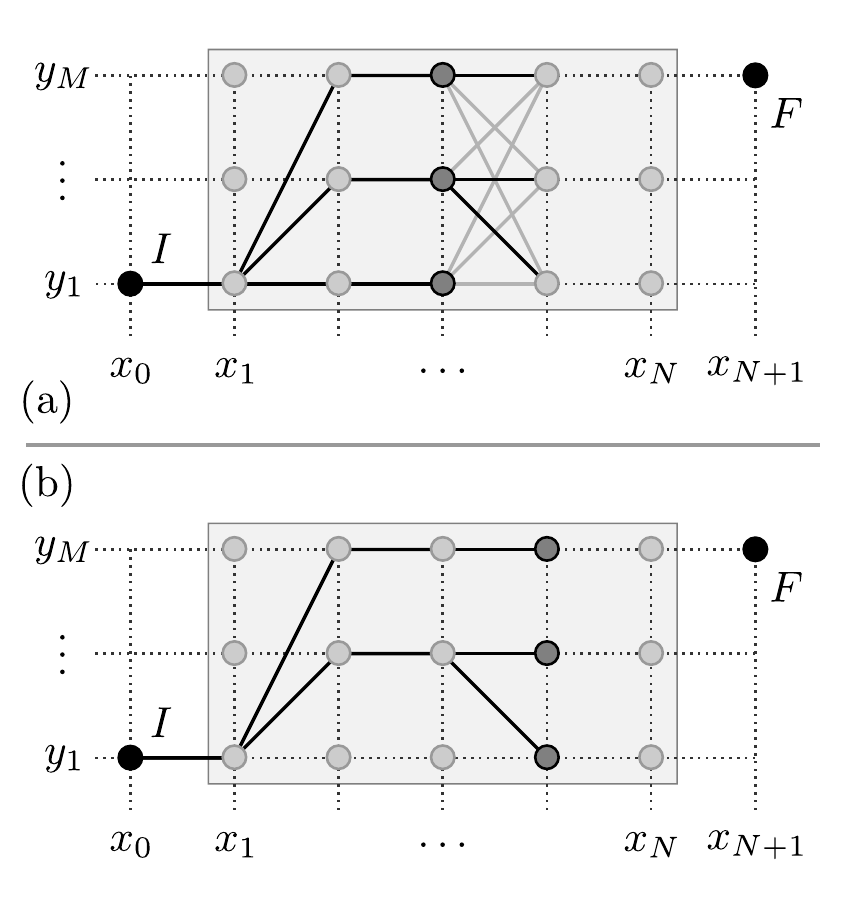}
\caption{\label{fig:DPM_1D}A depiction of the Dynamic Programming Method calculating the optimal path between the initial point $I$ and final point $F$ across the search space. (a) All successors are generated from the tip of each path at $x$ to points at $x + 1$, but for each different $y$-value, only the successor with the best partial score (indicated with solid lines) is kept. (b) The suboptimal successors have been pruned, keeping just one path for each $y$-value.}
\end{figure}

Given the initial and final points, the different values of $x$ within the search space can be enumerated as $x_0, x_1, \dots, x_N, x_{N+1}$ where $x_0 = x_i$ and $x_{N+1} = x_f$. Each step of any path is assumed to advance by one step in the $x$ direction, so a complete path must have $N$ steps from start to finish. The different $y$-values in the search space can be listed $y_1, y_2, \dots, y_M$, but $y_1$ and $y_M$ do not necessarily correspond to the initial and final $y$-values. The grid axes in Fig. \ref{fig:DPM_1D} are labelled according to these definitions. \\

A breadth-first search begins from the initial point with $x = x_0$. A successor path is generated for each possible value of $y$ for $x = x_1$. Then, for each of these paths, further successors are generated by considering all choices of $y$ for $x = x_2$, and so on, until all possible paths to the destination $(x_{N+1}, y_f)$ are calculated, and the path with the best score can be determined. This method is exhaustive and guaranteed to obtain the optimal path, but its memory and time usage scale exponentially with the dimensions of the search space, as all $M^N$ paths are stored and generated simultaneously. \\

The Dynamic Programming Method, illustrated in Fig. \ref{fig:DPM_1D}, improves on the BFS algorithm by assuming that only the most recent step in a path (the ``tip'') affects the action of subsequent steps. After successor paths are calculated for any $x > x_1$, there will be multiple paths with the same value of $y$ at their tip; comparing the partial action for each of these paths allows all but the minimal one to be discarded [Fig. \ref{fig:DPM_1D}(a)]. This removes the exponential memory and time requirements by only keeping a maximum of $M$ paths between each step [Fig. \ref{fig:DPM_1D}(b)]. The method is able to obtain the optimal path out of $M^N$ possibilities while only exploring $MN$ paths. \\

In order to adapt DPM to calculate fission paths traversing discontinuous regions of 2D and three-dimensional (3D) PESs rather than tunnelling through potential energy barriers, the following modifications are used in this work:
\begin{itemize}
\item The total energy of the path $E_\mathrm{tot}$ is minimised instead of the total action. Since the action is proportional to the path length, minimising it skews the optimisation towards shorter paths, even if they are energetically unfavourable. The total energy of a path may be defined as
\begin{equation}
E_\mathrm{tot}[\mathcal{P}] = \sum_{j=1}^N E(p_j), \label{eq:dpm_total_energy}
\end{equation}
where $p_j \equiv (x_j, y_j)$ is the $j$th element in the ordered set $\mathcal{P}$ representing the path. The energies of the initial and final points are neglected as they will be identical for all paths.
\item In order to restrict the search to continuous paths, a maximum gradient $\Delta_\mathrm{max}$ is imposed such that a step along the path in $x$ can only vary in $y$ (or any other dependent coordinate) by up to $\Delta_\mathrm{max}$ mesh steps.
\item As an additional condition, the calculated overlap [Eq. (\ref{eq:overlap})] between the wavefunctions of the tip of the path and its successors must exceed a threshold value $\mathcal{N}_\mathrm{min}$ in order for each successor to be considered at all.
\end{itemize}
Minimising the total energy instead of the total action also removes the need for the collective inertias to be calculated at each point on the grid. The two added conditions decrease the multiplicity of the search space by restricting possible successors. However, the values of the maximum gradient $\Delta_\mathrm{max}$ and overlap threshold $\mathcal{N}_\mathrm{min}$ must be chosen sensibly with respect to the situation, otherwise the algorithm may fail to find a complete path. \\

Recall that the HFB algorithm which produces the initial fission paths calculates energies under the adiabatic approximation, producing ground state solutions with the local minimum energy under the given constraints. Using DPM in the way described above relaxes the locality of this assumption to apply to a path instead of individual points: each point on the path may deviate from the local minimum energy if this lowers the total energy integrated over the path. When the gradient and overlap conditions are applied, the method produces fission paths which are continuous in both energy and nuclear collective coordinates, while retaining the physical reasoning of adiabaticity. \\

It should be noted that the assumption of adiabaticity becomes invalid as the nucleus approaches scission. This means DPM as formulated in this work is ill-suited to handle discontinuities near scission, including the scission line itself. Various methods have been proposed to handle non-adiabatic effects, such as time-dependent mean-field descriptions \cite{simenel2014,bulgac2016} or the use of additional PESs to account for quasiparticle excitations \cite{bernard2011}. The latter approach is particularly notable as DPM could simply be applied to the individual PESs.

\subsection{\label{sec:frontier_dpm}``Frontier DPM'' for 2D surface smoothing}

The definition of the system for finding a two-dimensional surface in a three-dimensional PES is a logical advancement from the case of the one-dimensional path. The search space is a three-dimensional grid in $(x,y,z)$ coordinates, where the values of $x$ run from $x_0$ to $x_{N+1}$, $y$ from $y_0$ to $y_{M+1}$, and $z$ from $z_1$ to $z_L$. The aim is to determine the values of $z$ for each of the $(x,y)$ coordinate pairs, which together define a surface covering the $x$-$y$ plane. Instead of initial and final points, enclosing boundary conditions must be specified to fix the $z$-values along the edges of the surface. \\

Performing a breadth-first search in three dimensions is truly a ``brute-force'' approach, and turns out to be impractical for all but the most trivial problems. The total number of possible surfaces is $L^{MN}$; the ``double exponential'' dependency scales extremely poorly with the size of the search space. It is necessary to find a more efficient method to proceed. \\

\begin{figure}
\includegraphics[width=8.6cm]{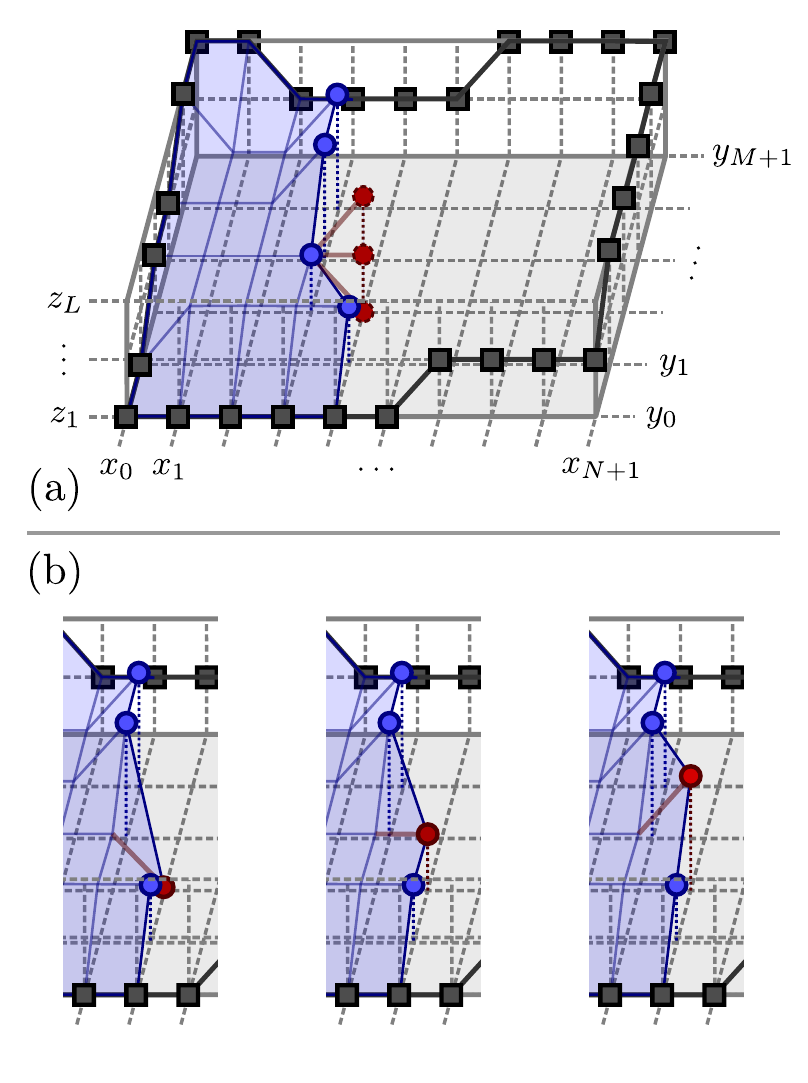}
\caption{\label{fig:DPM_2D}Illustrations of Frontier DPM calculating the optimal surface inside specified boundary conditions (black squares). (a) The current ``frontier'' of the surface is marked with blue circles, while the red circles with dashed outlines represent the different $z$-values for the next $(x,y)$ point to be added to the surface. (b) The frontiers of the successor surfaces are shown, with none having identical frontier configurations.}
\end{figure}

Since the Dynamic Programming Method drastically improves on the performance of the BFS in one dimension, generalising the method to two dimensions is a natural next step. To do so, first the following concepts are defined, with reference to the illustration in Fig. \ref{fig:DPM_2D}(a):
\begin{itemize}
\item The boundary (black squares) is the set of $(x,y,z)$ points corresponding to the boundary conditions, enclosing the surface to be calculated.
\item A partial surface (blue shaded area) is a set of $(x,y,z)$ points within the boundaries selected by the algorithm plus the boundary points.
\item The frontier of a surface (blue circles) is the subset of points within the partial surface, not including any boundary points, which are adjacent in $(x,y)$ to points not yet added to the surface.
\item For a given $(x,y)$ adjacent to the frontier (or the boundary, if the frontier is empty), a partial surface yields a successor for each possible $z$-value for the chosen $(x,y)$ coordinates (red circles).
\end{itemize}

The core iteration of DPM in one dimension is to generate successors from the current list of paths for a particular value of $x$, then compare the partial paths with the same $y$-value ``tip'', and keep only the best one. The two-dimensional analogue is to generate successors from the current list of surfaces for a particular $(x,y)$, then compare the partial surfaces with the same frontier, and keep only the best one. \\

Although a given partial surface will always generate successors with different frontiers as illustrated in Fig. \ref{fig:DPM_2D}(b), multiple surfaces may produce successors with the same frontier; when this occurs, only the lowest-energy surface is kept. It can be estimated that up to $L^M$ surfaces (or $L^N$, if $N < M$) will be retained between steps of the calculation, with a total number of surfaces on the order of $NML^{M+1}$ being considered to find the optimal complete surface. When compared to the BFS, the ``double exponential'' dependency has been removed, which means that the time and memory requirements of ``Frontier DPM'' are much more reasonable. \\

The Frontier DPM algorithm can be expressed in the following sequence of steps:
\begin{enumerate}
\item The initial set of partial surfaces contains only one surface consisting of the boundary points.
\item Choose the next point in the $(x,y)$ plane to add to the surface, preferring the point adjacent to the largest number of non-boundary points already in the surface, and then the point adjacent to the most boundary points [see Fig. \ref{fig:DPM_2D}(a)].
\item For each possible $z$-value at this point, generate a successor surface from each existing surface by connecting the point [see Fig. \ref{fig:DPM_2D}(b)]. Determine the set of $(x,y,z)$ points forming the frontier for the new surface; if any successors generated in this step have identical frontiers, discard the one with higher total energy.
\item Repeat from step 2 until the entire surface within the boundaries is filled.
\end{enumerate}
As with the adaptation of the original DPM from fission lifetime calculations to fission path optimisation, maximum gradient (in the $z$ direction) $\Delta_\mathrm{max}$ and minimum overlap conditions $\mathcal{N}_\mathrm{min}$ are imposed during step 3 to force solutions to be continuous in the collective coordinates.

\section{\label{sec:results}Results and discussion}
\subsection{Smoothing 1D discontinuities in $\elem{252}{}{Cf}$}

\begin{figure}
\includegraphics[width=8.6cm]{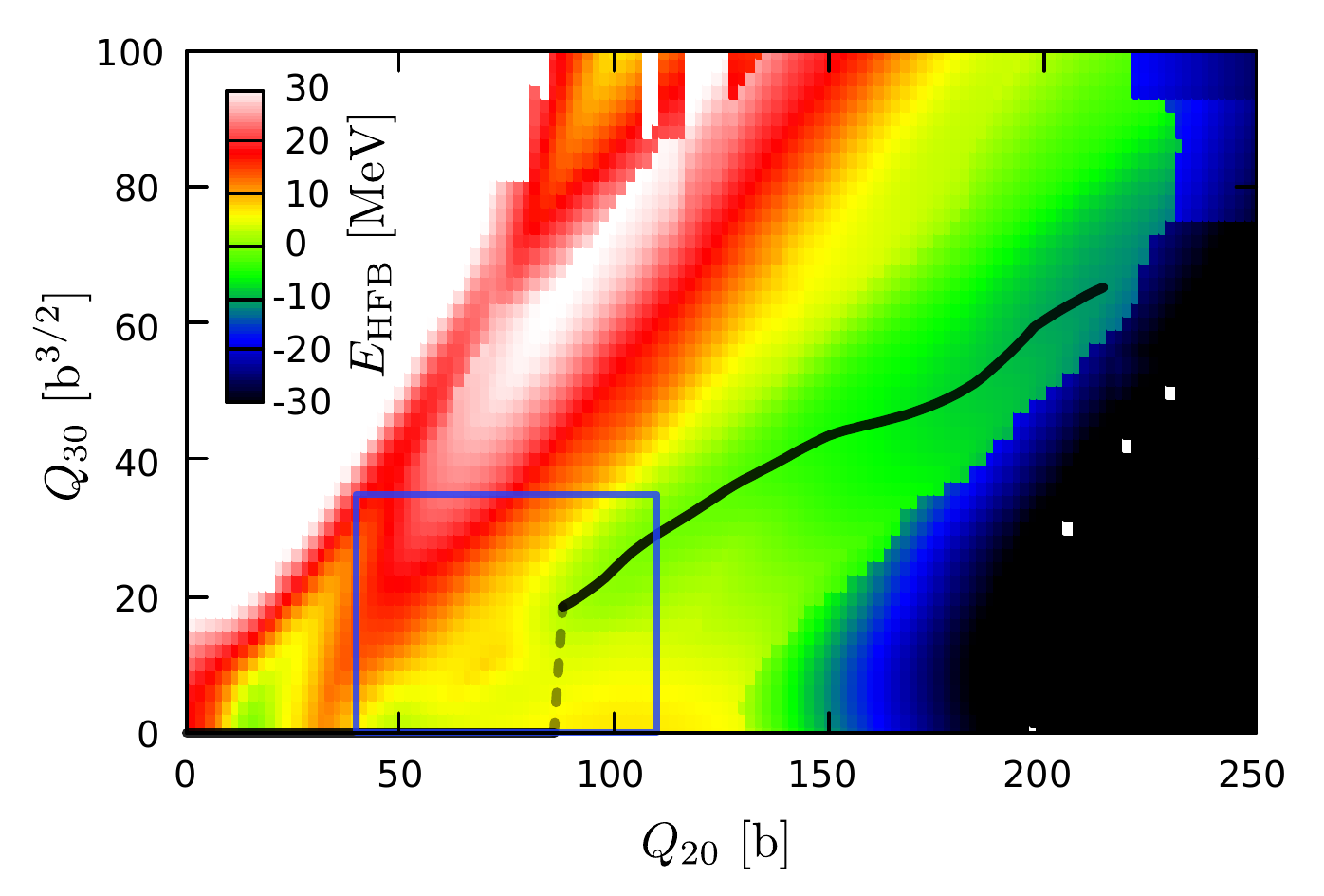}
\caption{\label{fig:252Cf_pes}The two-dimensional PES of $\elem{252}{}{Cf}$ in $Q_{20}$ and $Q_{30}$, generated with self-consistent HFB using the D1S Gogny interaction. $E_\mathrm{HFB}$ is calculated relative to the ground state energy of the nucleus. The 1D fission path is plotted with a black line, with the dashed region representing the discontinuous transition from symmetric to asymmetric nuclear shape. The blue rectangle indicates the region of interest analysed in subsequent figures.}
\end{figure}

The PES of $\elem{252}{}{Cf}$ was selected as a first candidate for testing the applicability of the Dynamic Programming Method to the smoothing of discontinuities in 1D fission paths. HFB solutions were calculated using an oscillator basis of $N_\perp = 14,\ N_z = 21$ shells, and \textsc{HFBaxial} was configured to vary the oscillator lengths. There is a clear one-dimensional discontinuity in the fission path where it transitions from symmetric to asymmetric configurations, shown in Fig. \ref{fig:252Cf_pes}.  After crossing the first fission barrier, the  fission path jumps suddenly from a symmetric configuration into the opening asymmetric fission valley at $Q_{20} \approx 88 \unitsn{b}$.

\subsubsection{Discontinuity analysis}

\begin{figure}
\includegraphics[width=8.6cm]{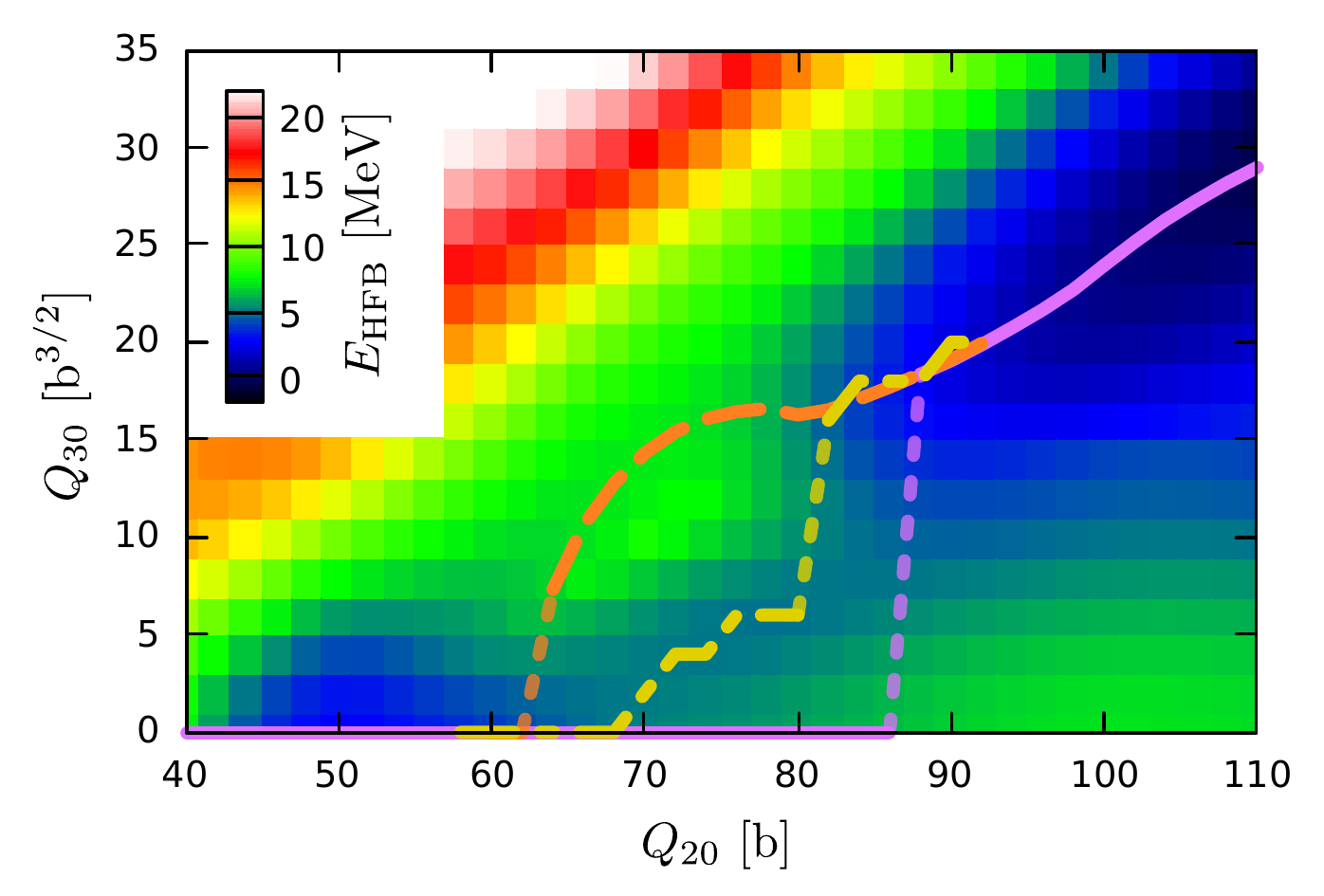}
\caption{\label{fig:252Cf_pes_zoomed}A zoomed-in view of the region marked in Fig. \ref{fig:252Cf_pes}. The HFB forward-propagated fission path is plotted in solid pink, while the path calculated in the reverse direction is shown with an orange dashed line. The adiabatic path derived from the 2D PES is drawn with a dash-dotted yellow line. The fainter dotted segments in each path indicate discontinuous jumps in $Q_{30}$.}
\end{figure}

By only constraining $Q_{20}$, the 1D fission path is essentially free to follow local minima without maintaining any continuity in other coordinates such as $Q_{30}$. Such paths are illustrated in Fig. \ref{fig:252Cf_pes_zoomed}, which depicts a zoomed-in region of the PES centred on the discontinuity. The fission path (the solid pink line) obtained by propagating the solution wavefunction in the increasing $Q_{20}$ direction remains at $Q_{30} = 0 \unitsn{b^{3/2}}$, even after the symmetric fission valley disappears at $Q_{20} \approx 70 \unitsn{b}$, then transitions suddenly to an asymmetric configuration with $Q_{30} \approx 18 \unitsn{b^{3/2}}$ at $Q_{20} \approx 88 \unitsn{b}$. \\

Further insight into the discontinuity can be obtained by generating the ``reverse'' fission path (the dashed orange line in the same figure), starting from the asymmetric fission valley and propagating the solution wavefunction in the decreasing $Q_{20}$ direction instead. This path remains in the asymmetric fission valley as long as possible until it vanishes, suddenly dropping to $Q_{30} = 0 \unitsn{b^{3/2}}$ at $Q_{20} \approx 62 \unitsn{b}$. The presence of multiple HFB solutions for the 1D fission path suggests that this discontinuity arises at the intersection between two fission valleys separated in $Q_{30}$; when fission paths are calculated only constraining $Q_{20}$, solutions jump suddenly between the two valleys in order to minimise their energy. \\

A closer look at the 2D PES in this region shows that there is a low-energy valley between the two HFB paths, raising the question: why does the forward fission path not follow this third valley? A likely cause is the fact that $Q_{30} = 0$ is a stationary point for any fixed $Q_{20}$ due to reflection symmetry. However, following local minima to construct the 1D fission path does not generally guarantee that the lowest energy path will be obtained. As a result, an additional adiabatic fission path (dash-dotted yellow line in Fig. \ref{fig:252Cf_pes_zoomed}) was calculated in the region by selecting values of $Q_{30}$ from the 2D PES which minimise the energy for each step in $Q_{20}$. The adiabatic path follows the low-energy valley as hoped, but it still contains a discontinuity in $Q_{30}$ at $Q_{20} \approx 80 \unitsn{b}$. It is reasonable to predict that a smoothed, continuous fission path will follow the adiabatic path closely up until this discontinuity. \\

\begin{figure}
\includegraphics[width=8.6cm]{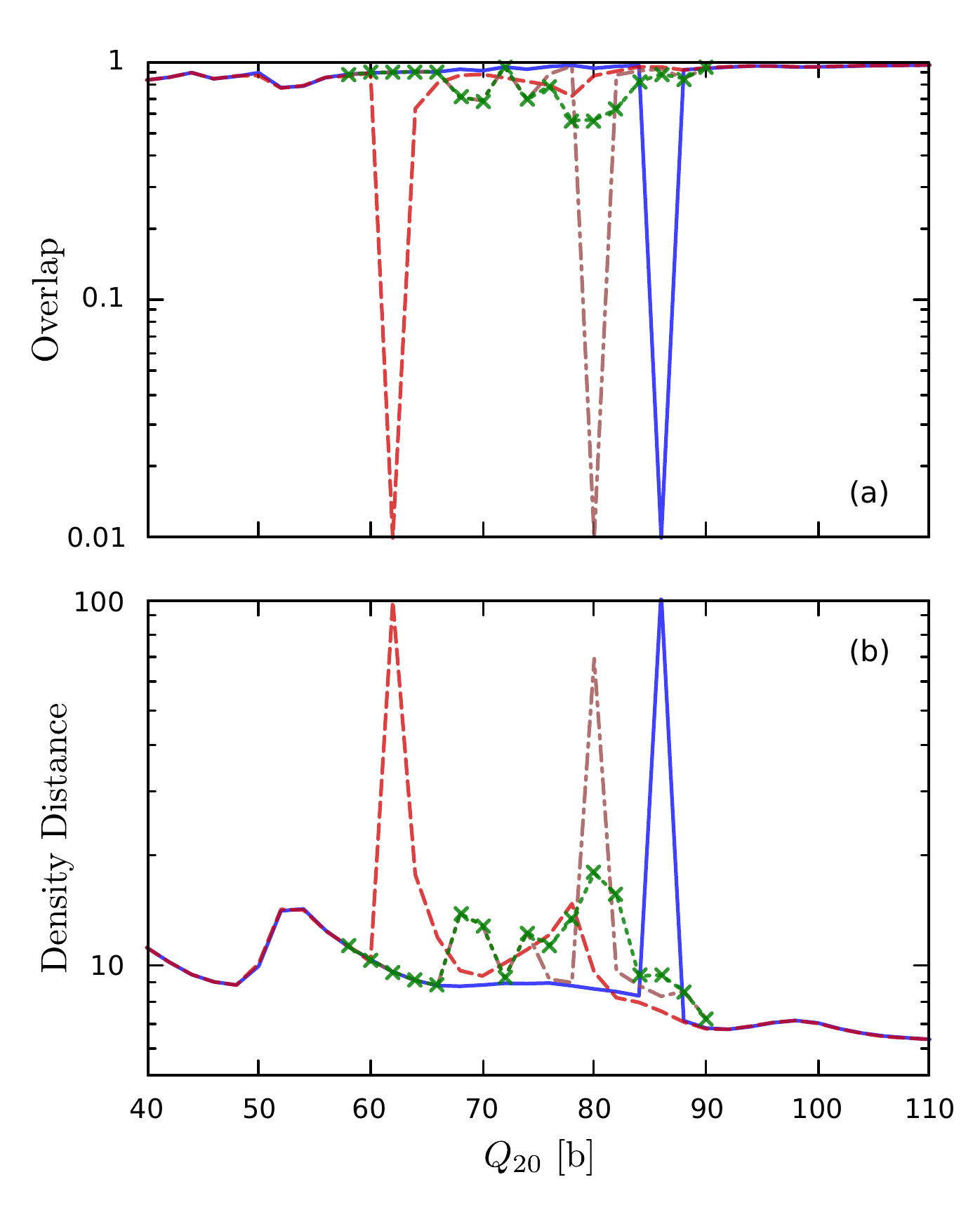}
\caption{\label{fig:252Cf_over_dd}Separate plots of the (a) overlaps and (b) density distances between adjacent points on the various fission paths, as functions of $Q_{20}$ over the same region as Fig. \ref{fig:252Cf_pes_zoomed}, with logarithmically-scaled $y$-axes. The HFB forward fission path is plotted in solid blue, the HFB reverse path in dashed red, and the derived adiabatic path in dash-dotted brown, with fainter dotted segments indicating discontinuities. The smoothed path calculated with the DPM method is plotted in dashed green with cross markers. The overlap approaches zero and the density distance becomes large across large changes in nuclear configuration, corresponding to the discontinuities in $Q_{30}$ in the forward and reverse HFB and the adiabatic fission paths. Note that overlaps below $0.01$ (including zero) in (a) are plotted as 0.01.}
\end{figure}

The overlaps (\ref{eq:overlap}) and density distances (\ref{eq:dens_dist}) can be calculated for the neighbouring points on the HFB fission paths and the adiabatic path. The results, plotted in Fig. \ref{fig:252Cf_over_dd}, are as expected: the overlap drops to zero and the density distance increases approximately tenfold when crossing between the two fission valleys in either direction. In order for the smoothing algorithm to be successful, it is expected that both of these metrics should be vastly improved by ``flattening'' the sharp changes in $Q_{30}$ over a larger region in $Q_{20}$.

\begin{figure}
\includegraphics[width=8.6cm]{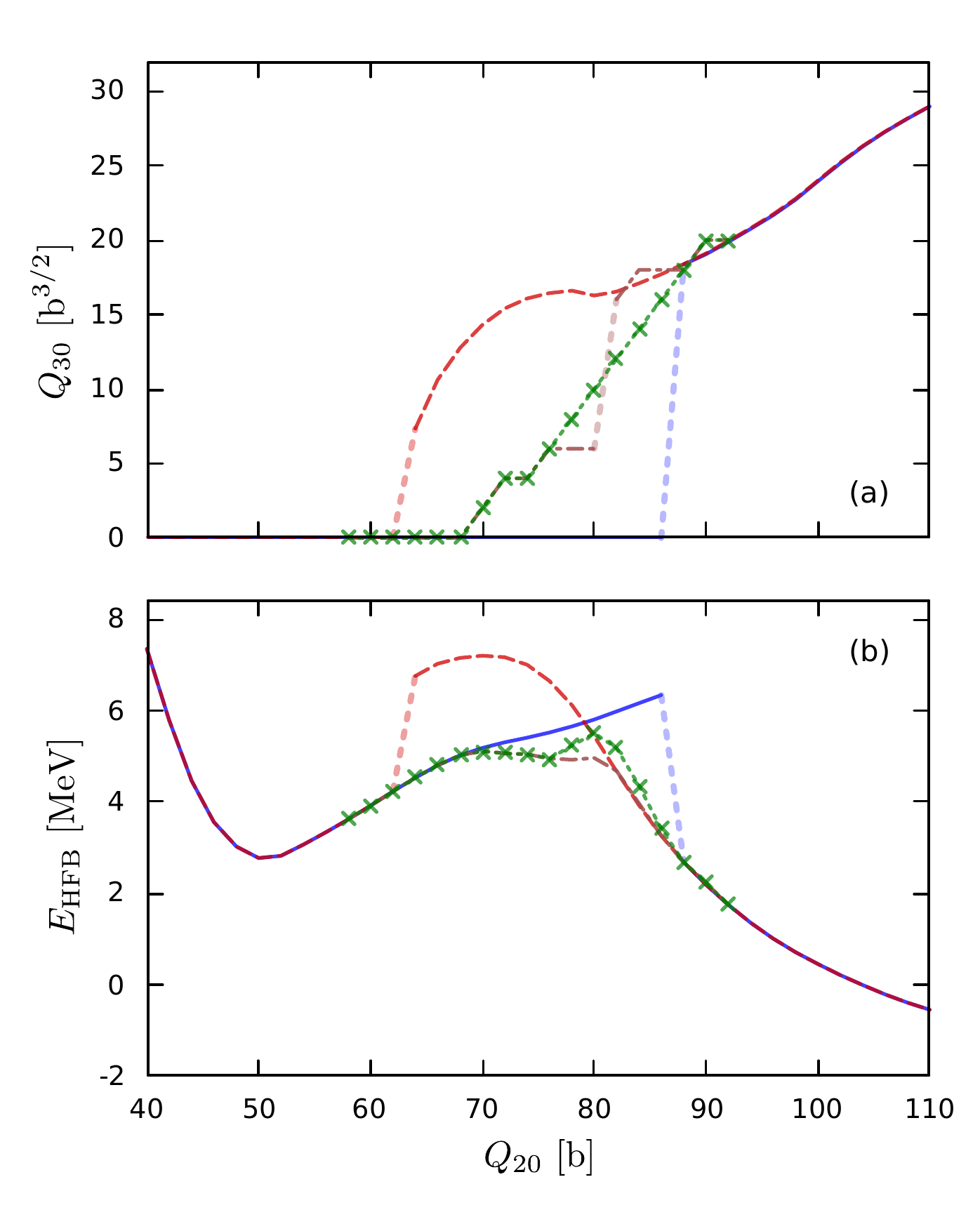}
\caption{\label{fig:252Cf_asym_paths}Plots comparing (a) $Q_{30}$ and (b) $E_\mathrm{HFB}$ relative to the ground state for the HFB forward- and reverse-propagated fission paths, adiabatic path and DPM smoothed path as functions of $Q_{20}$ over the region marked in Fig. \ref{fig:252Cf_pes}. The different fission paths are drawn with the same styles as in Fig. \ref{fig:252Cf_over_dd}.}
\end{figure}

\subsubsection{Parameters}

Since the identified discontinuity is in $Q_{30}$, all that is needed to generate a smoothed 1D fission path is the 2D $Q_{20}$-$Q_{30}$ PES in the discontinuous region. In this case, the region of the PES used for smoothing was within the ranges $Q_{20} \in [58, 92] \unitsn{b}$ and $Q_{30} \in [0, 20] \unitsn{b^{3/2}}$. The desired path in this region was specified by the initial and final points $(58, 0)$ and $(92, 20)$ respectively, expressed as $(Q_{20}$, $Q_{30}$) in barn powers. The maximum permitted gradient was $\Delta_\mathrm{max} = 1$ step ($2 \unitsn{b^{3/2}}$) in $Q_{30}$ per step ($2 \unitsn{b}$) in $Q_{20}$, and the minimum allowed overlap was $\mathcal{N}_\mathrm{min} = 0.3$.

\subsubsection{Results}

The lowest-energy smoothed path from the DPM is plotted with green dashed lines and crosses over the HFB paths (following local minima forwards or backwards) and adiabatic paths (following global minima) in Fig. \ref{fig:252Cf_asym_paths}. As expected, the DPM path begins by following the adiabatic path closely, but increases in $Q_{30}$ at a steady rate so that the nucleus deforms continuously in $Q_{20}$ and $Q_{30}$ along the path. This results in a slightly higher second barrier than that suggested by the adiabatic path. This outcome is quite reasonable given that the adiabatic path entirely ignores the energy cost of travelling between the symmetric and asymmetric fission valleys; it would be worthwhile to investigate the effects of this difference on the predicted spontaneous fission lifetimes and fragment charge-mass distributions of the nucleus. \\

While the gradient and overlap restrictions must be imposed to obtain continuous paths, it is also evident that they restrict the space of possible solutions. In particular, the maximum allowed gradient must be chosen sensibly with respect to the initial and final points imposed for the path; if it is too small, the algorithm will be unable to consider paths except for one with a constant gradient, or it may not be able to find a solution at all. The overlap threshold is less sensitive, but it must nevertheless be set depending on the mesh size of the search space to allow ample variations in the nuclear state between steps. \\

The overlaps and density distances between neighbouring points on the smoothed path are plotted over those of the other paths in Fig. \ref{fig:252Cf_over_dd}. As desired, the adapted DPM algorithm has smoothed the discontinuities according to both metrics, removing the regions of zero overlap and high density distance present in the previously calculated fission paths. Between the locations of the previous discontinuities, the overlaps are slightly lowered and density distances slightly elevated; this occurs because the nuclear configuration now changes in $Q_{30}$ more consistently as the path travels continuously between the fission valleys, rather than all at once across a discontinuity.

\subsection{Comparing smoothed 1D paths in $\elem{252}{}{Cf}$ and $\elem{222}{}{Th}$}

\begin{figure}
\includegraphics[width=8.6cm]{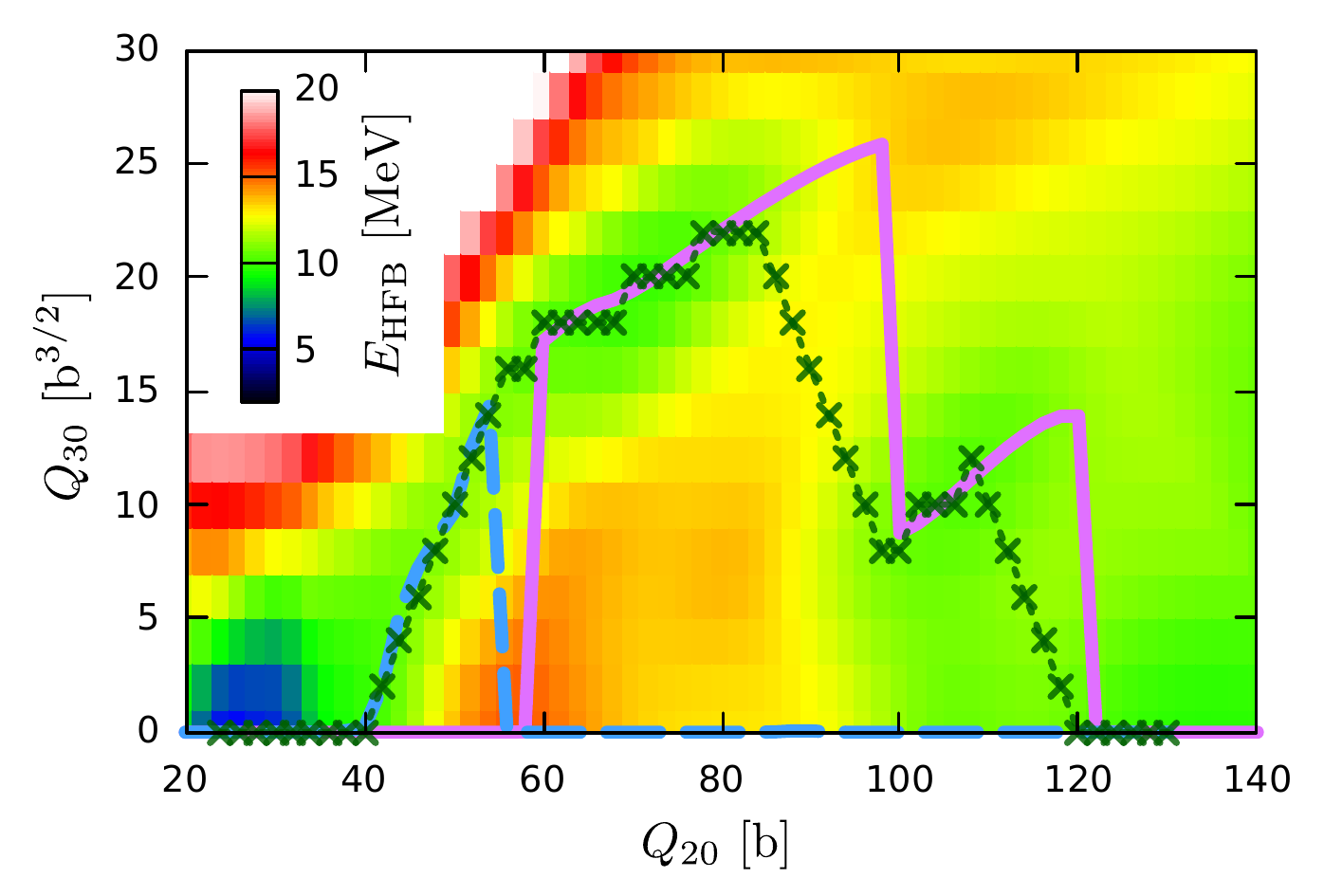}
\caption{\label{fig:222Th_pes_zoomed_paths}A subsection of the 2D PES in $Q_{20}$ and $Q_{30}$ for $\elem{222}{}{Th}$, generated with $N_\perp = 15,\ N_z = 22$ shells and fixed oscillator lengths $b_\perp = 2.0,\ b_z = 3.0$. Energies are shown relative to the ground state. The HFB paths propagated in the $+Q_{20}$ and $-Q_{20}$ directions are drawn with solid pink and dashed blue, respectively, while the smoothed path generated using DPM is depicted in dotted green with crosses.}
\end{figure}

As an additional evaluation of DPM's usefulness, the fission paths it generated were compared to those produced with the adiabatic and linear interpolation methods outlined in Sec. \ref{sec:smoothing_discontinuities}. These comparisons were performed on 1D discontinuities in $\elem{}{252}{Cf}$ and $\elem{}{222}{Th}$. The discontinuity in $\elem{}{252}{Cf}$ is analyzed in the previous section, while the relevant portion of the 2D PES for $\elem{}{222}{Th}$ is presented in Fig. \ref{fig:222Th_pes_zoomed_paths}. The transition from the initial potential well to asymmetry before returning to the symmetric fission valley is poorly represented by the HFB paths (solid pink, dashed blue), but is captured more accurately by the DPM path (dotted green with crosses). \\

\begin{figure}
\includegraphics[width=8.6cm]{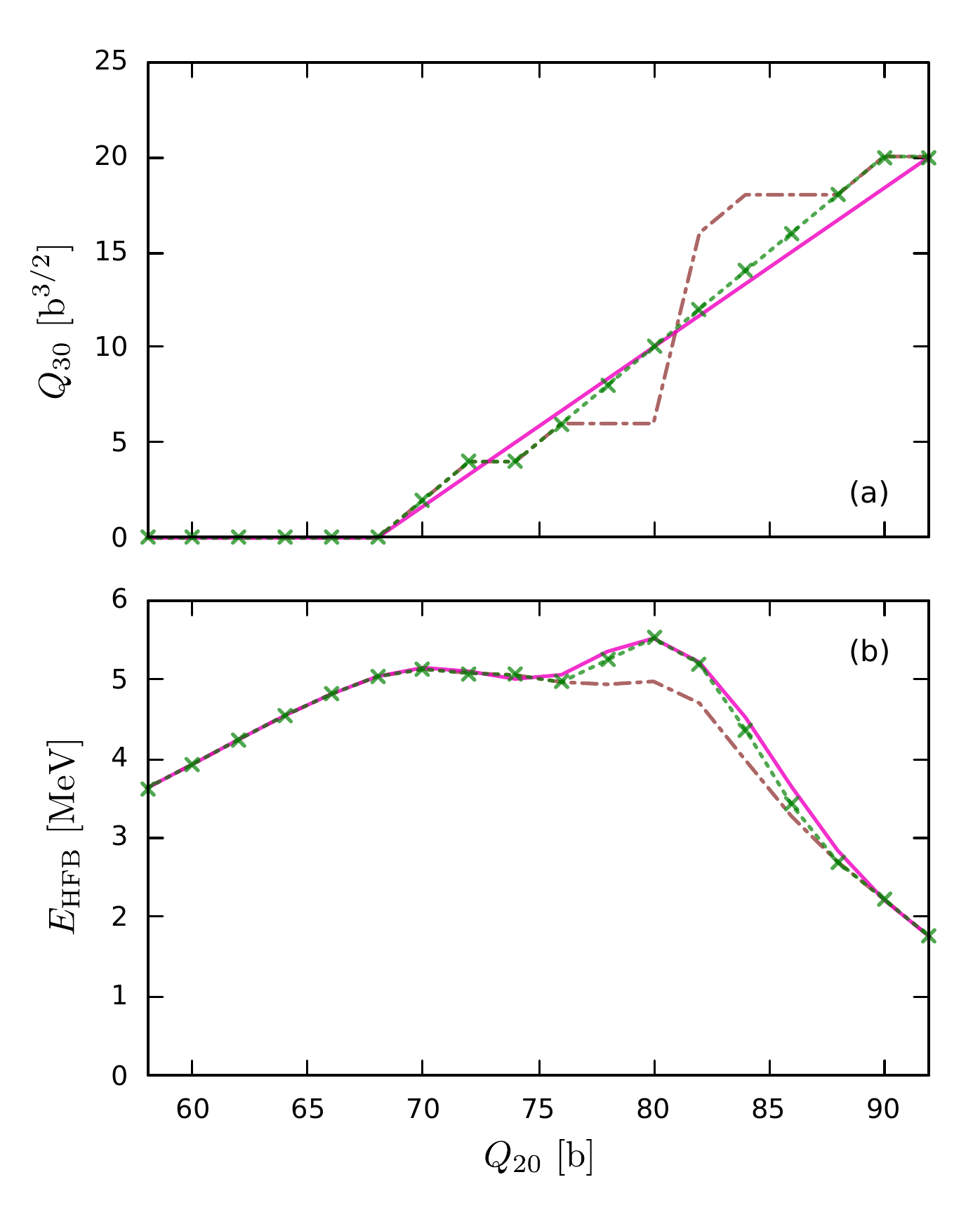}
\caption{\label{fig:252Cf_ada_lin_q30_ener}Plots of (a) $Q_{30}$ and (b) HFB potential energy relative to the ground state as functions of $Q_{20}$, comparing smoothed fission paths calculated with different methods over the 1D discontinuity $\elem{252}{}{Cf}$. The adiabatic path is drawn in dash-dotted brown, the linear path in solid pink, and the DPM path in dotted green with crosses.}
\end{figure}

\begin{figure}
\includegraphics[width=8.6cm]{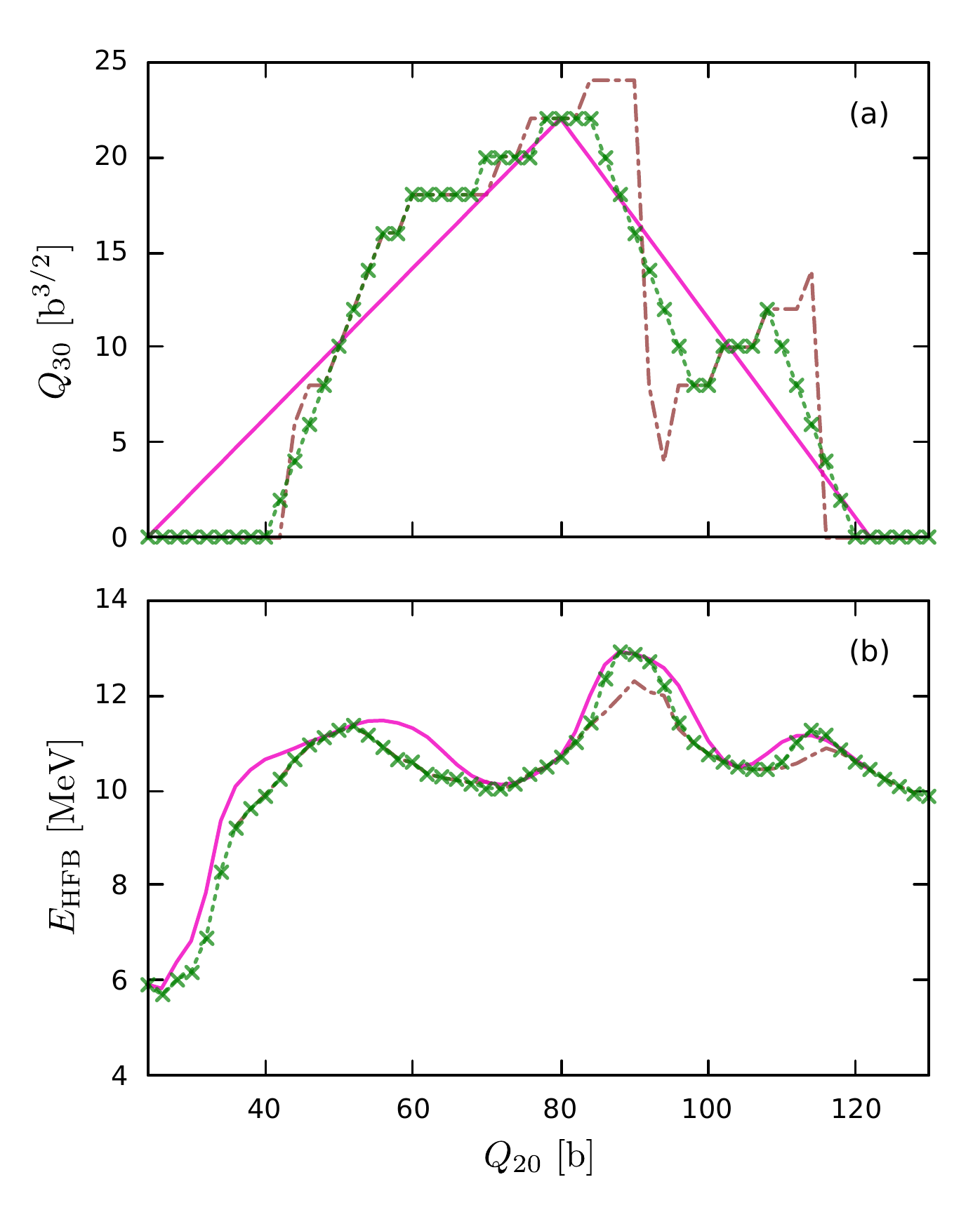}
\caption{\label{fig:222Th_ada_lin_q30_ener}Identical plots to Fig. \ref{fig:252Cf_ada_lin_q30_ener} but for $\elem{222}{}{Th}$. The fission paths are drawn with the same styles.}
\end{figure}

Linear paths were generated for each PES by choosing suitable vertices by eye so that the straight lines connecting them best followed the low-energy valleys on the 2D PES. The selected vertices for each nuclide in $(Q_{20}\,\mathrm{[b]}, Q_{30}\,\mathrm{[b^{3/2}]})$ notation were:
\begin{itemize}
\item $\elem{252}{}{Cf}$: $(58,0)$ to $(92,20)$
\item $\elem{222}{}{Th}$: $(24,0)$ to $(80,22)$ to $(122,0)$ (two segments)
\end{itemize}
\textsc{HFBaxial} was used to calculate these new paths, interpolating the value of $Q_{30}$ between the vertices. Plots of the adiabatic, linear, and DPM paths are shown in Fig. \ref{fig:252Cf_ada_lin_q30_ener} for $\elem{252}{}{Cf}$ and in Fig. \ref{fig:222Th_ada_lin_q30_ener} for $\elem{222}{}{Th}$. \\

In both cases, the paths smoothed with either DPM or linear interpolation display increased barrier heights compared to the paths generated with the adiabatic method. While the DPM and linear paths appear very similar over the discontinuity in $\elem{252}{}{Cf}$, this is not true for $\elem{222}{}{Th}$. In the latter case, the linear path deviates significantly in $Q_{30}$ from the DPM and adiabatic paths: although the linear and DPM paths overlap at the peaks of the barriers, the energies of the linear path are up to 1 MeV higher in a wide range either side of the barriers. These comparisons show that smoothed paths constructed via linear interpolation are only of similar quality to those produced using DPM when the path between fission valleys happens to be well described by a straight line. In PESs such as for $\elem{222}{}{Th}$ where the transition is more complex in shape, the linear fitting method performs more poorly. Its accuracy also depends on the choice of interpolation endpoints, and hence is subject to the variation of human interpretation. \\

\begin{figure}
\includegraphics[width=8.6cm]{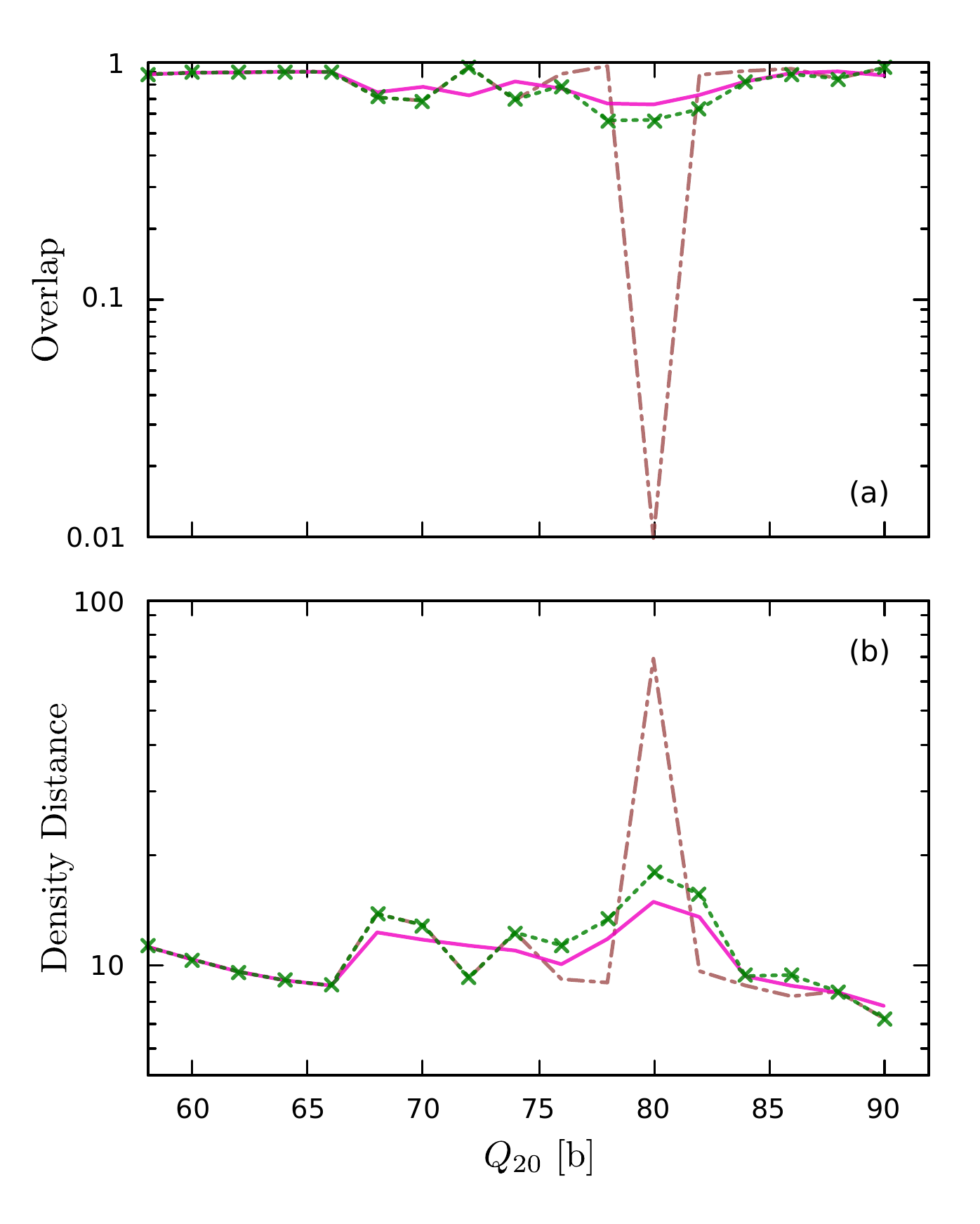}
\caption{\label{fig:252Cf_ada_lin_over_dd}Plots of the (a) overlaps and (b) density distances between adjacent points along different fission paths for $\elem{252}{}{Cf}$ smoothed across a 1D discontinuity. The adiabatic, linear, and DPM paths are shown using the same styles as the immediately preceding figures. Overlaps below 0.01 (including zero) in (a) are plotted as 0.01.}
\end{figure}

\begin{figure}
\includegraphics[width=8.6cm]{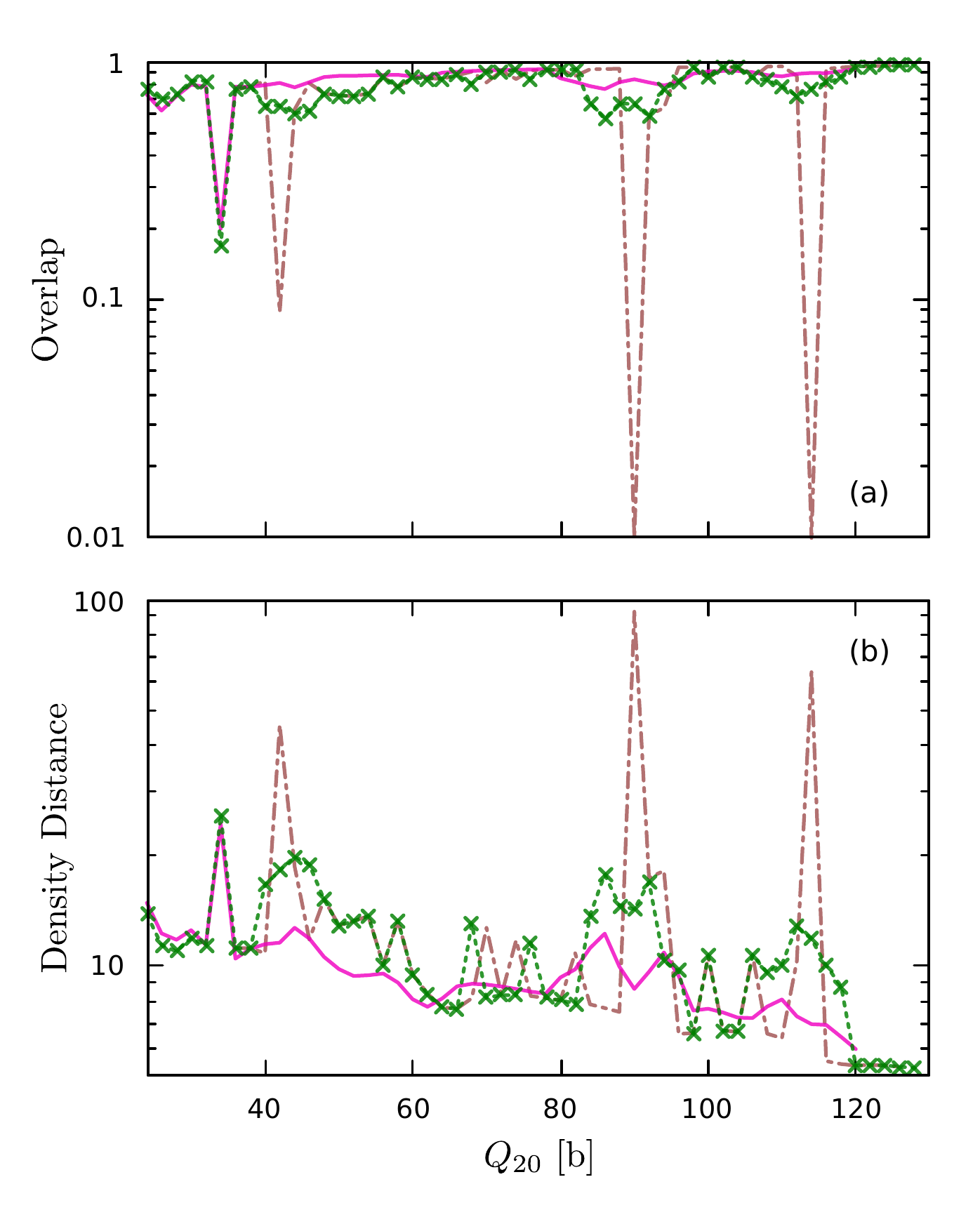}
\caption{\label{fig:222Th_ada_lin_over_dd}The same graphs as Fig. \ref{fig:252Cf_ada_lin_over_dd}, but showing results for $\elem{222}{}{Th}$. The fission paths are drawn with the same styles.}
\end{figure}

The overlaps and density distances between points on the various paths were also calculated to examine their smoothness. These quantities are displayed in Fig. \ref{fig:252Cf_ada_lin_over_dd} for $\elem{252}{}{Cf}$ and in Fig. \ref{fig:222Th_ada_lin_over_dd} for $\elem{222}{}{Th}$. While the adiabatic paths contain low overlaps and high density distances as expected, the linear paths appear to perform better than the DPM paths, with slightly increased overlaps and lowered density distances on average. However, the nature of the linear interpolation method means that its paths were not restricted to discrete values of $Q_{30}$ in $2 \unitsn{b^{3/2}}$ intervals, as was the case with DPM. Furthermore, the linear interpolation method prioritises smoothness in the changes of the nuclear configuration without any regard for the physical argument of adiabaticity. In contrast, DPM is formulated to balance smoothness of coordinates with minimisation of the overall energy. From this perspective, it is unsurprising that the linear interpolation method produces slightly smoother paths compared to DPM, as the requirement for continuity does not compete with other constraints. \\

In summary, the method of constructing linearly interpolated paths only is comparable to DPM in accuracy and smoothness when applied to simple 1D discontinuities whose crossings are well approximated by straight lines in the collective coordinate space. Even when both methods can produce similar results, the Dynamic Programming Method should be preferred due to stronger physical motivations of smoothness as well as adiabaticity. The requirement of human input for the linear interpolation method also makes it impractical to apply to higher dimensions, while DPM has been generalised to the smoothing of 2D surfaces in the form of Frontier DPM.

\subsection{Smoothing 2D discontinuities in $\elem{218}{}{Ra}$}

To test the effectiveness of the newly-devised Frontier DPM method, the 2D PES of $\elem{218}{}{Ra}$ was selected. It was discovered that discontinuities in the initial PES calculated with a harmonic oscillator basis of $N_\perp = 14,\ N_z = 21$ shells were spuriously caused by large variations of oscillator lengths between states, as described in Sec. \ref{sec:background_discontinuities_numerical}. To prevent these fake discontinuities from arising, the PES was recalculated with the oscillator lengths fixed as $b_\perp = 2.0,\ b_z = 3.1$ without dynamic adjustment. The size of the harmonic oscillator basis was increased to $N_\perp = 15,\ N_z = 22$ to compensate for any loss of precision due to these constraints. \\

\begin{figure}
\includegraphics[width=8.6cm]{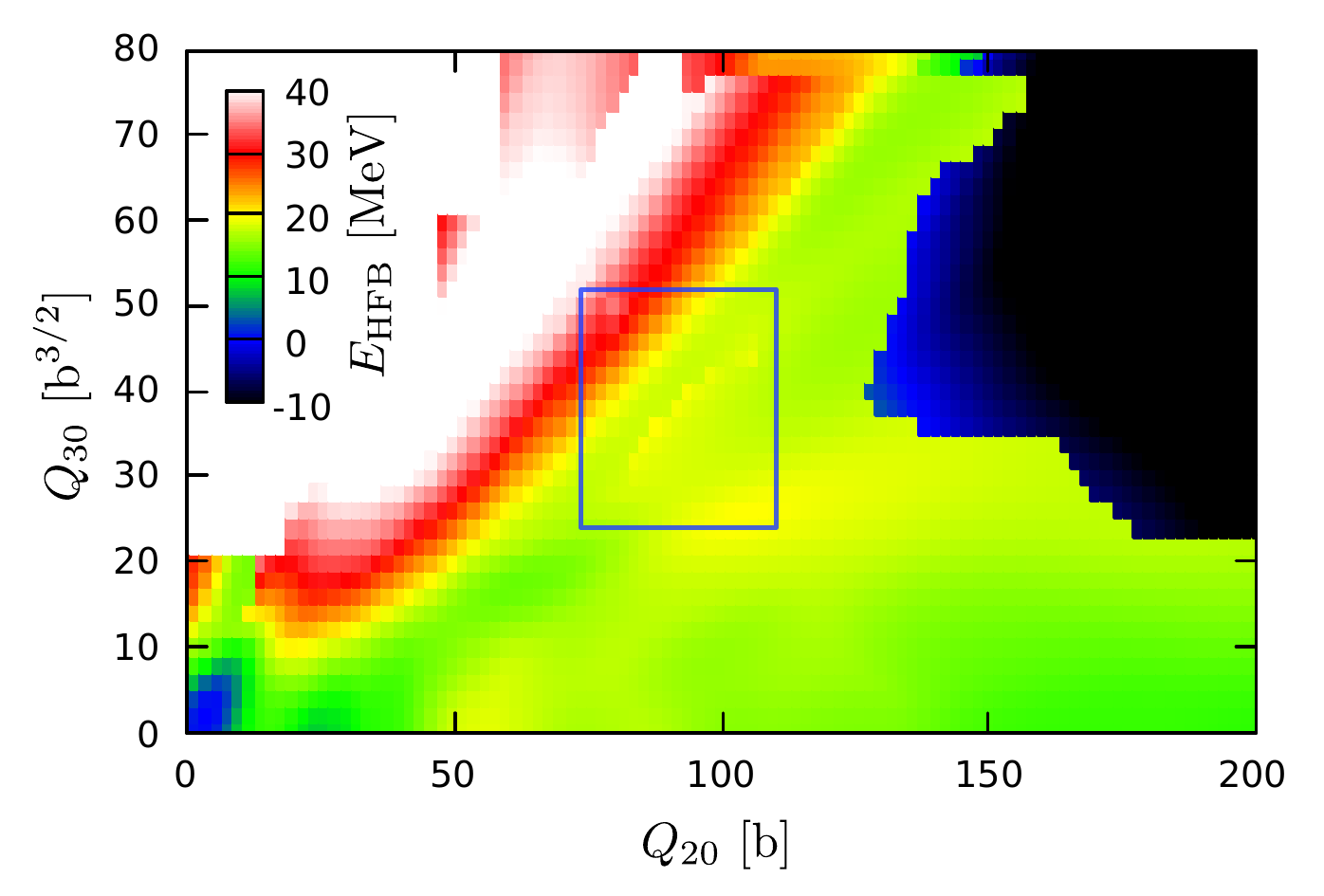}
\caption{\label{fig:218Ra_pes}The two-dimensional PES of $\elem{218}{}{Ra}$ in $Q_{20}$ and $Q_{30}$, generated with self-consistent HFB using the D1S Gogny interaction. $E_\mathrm{HFB}$ is shown relative to the ground state energy of the nucleus. The blue rectangle indicates the region of interest analysed in subsequent figures.}
\end{figure}

\begin{figure}
\includegraphics[width=8.6cm]{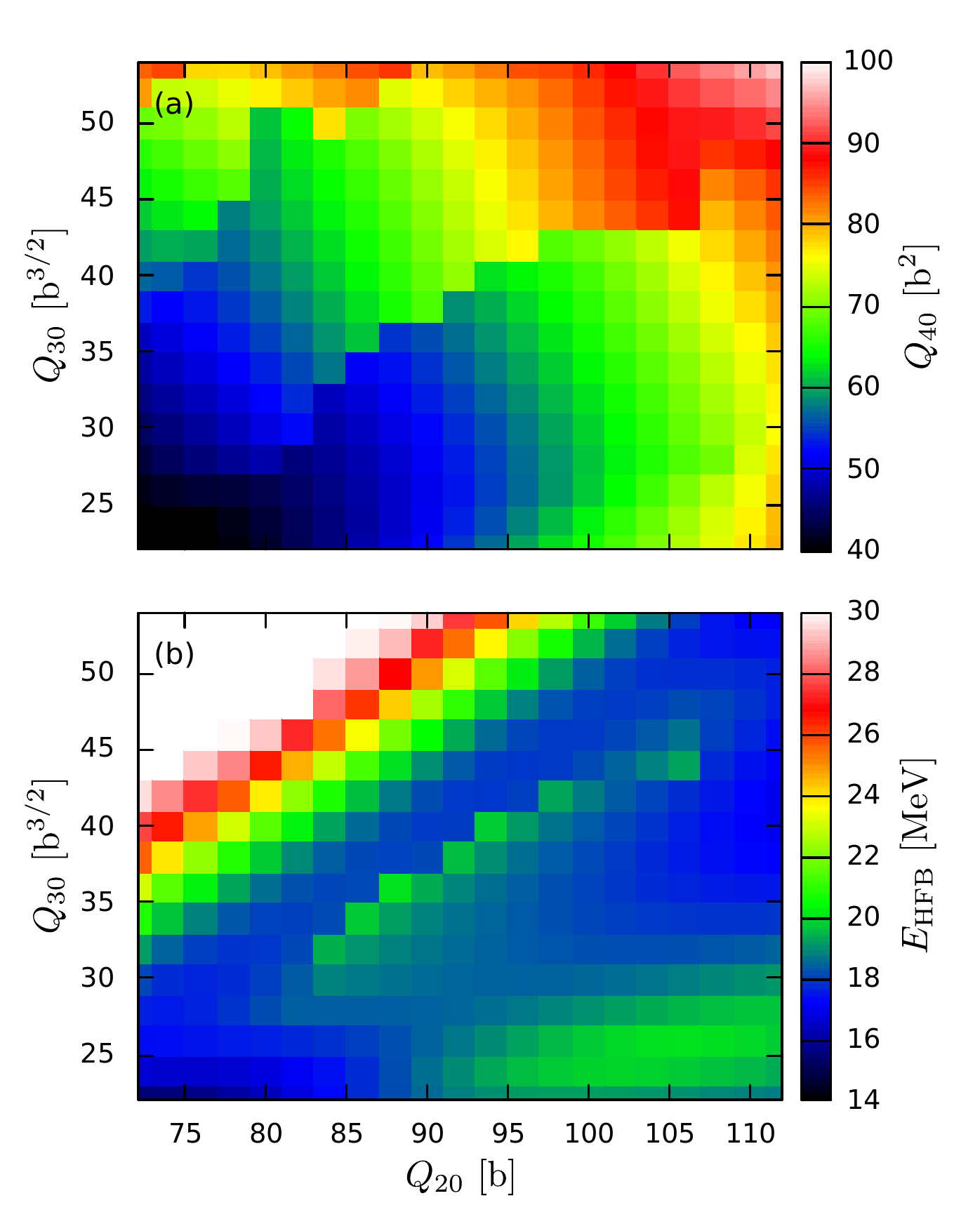}
\caption{\label{fig:218Ra_q40_ener}Plots of (a) $Q_{40}$ and (b) HFB potential energy relative to the ground state as functions of $Q_{20}$ and $Q_{30}$ over the region indicated in Fig. \ref{fig:218Ra_pes}.}
\end{figure}

The full PES is shown in Fig. \ref{fig:218Ra_pes}. A two-dimensional discontinuity occurs in the $Q_{40}$ coordinate, forming a diagonal shape bounded approximately by $Q_{20} \in [74, 110] \unitsn{b}$ and $Q_{30} \in [24, 52] \unitsn{b^{3/2}}$. It is identifiable by a sharp jump of around $10 \unitsn{b^2}$ in the $Q_{40}$ direction, indicative of a transition in the HFB solution between two competing fission valleys. The discontinuity in $Q_{40}$ is accompanied by a shift of 2-3$\unitsn{MeV}$ in potential energy. Magnified plots of $Q_{40}$ and $E_\mathrm{HFB}$ in the region are shown in Fig. \ref{fig:218Ra_q40_ener}. The sharp diagonal ``ridges'' in the centre of both plots indicate the presence of a discontinuity.

\subsubsection{Discontinuity analysis}

\begin{figure}
\includegraphics[width=8.6cm]{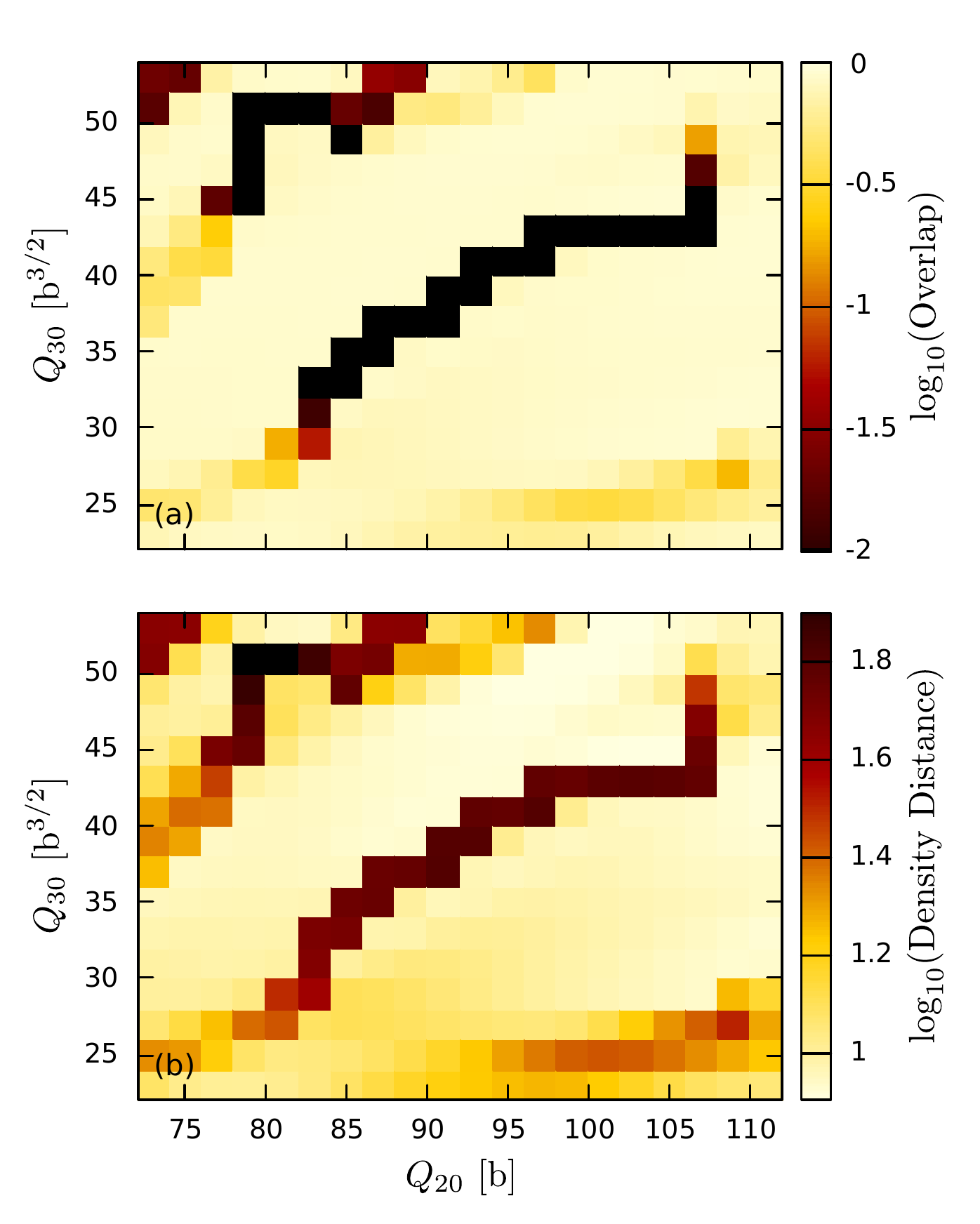}
\caption{\label{fig:218Ra_over_dd}Logarithmic plots of the (a) overlap and (b) density distance between neighbouring points as functions of $Q_{20}$ and $Q_{30}$ over the region indicated in Fig. \ref{fig:218Ra_pes}. In (a), logarithms of overlaps smaller than $-2$ (including $-\infty$ for overlaps equal to zero) are denoted with solid black. The darkest areas represent low overlaps in (a) and high density distances in (b), corresponding to large changes in nuclear configuration in these regions.}
\end{figure}

Deeper analysis of the nuclear structure and dynamics of $\elem{218}{}{Ra}$ would be required to fully understand the origins of the discontinuity, but in the present work it is sufficient to suppose that there are multiple overlapping fission valleys in the region, and to examine the overlaps and density distances to confirm the presence of the discontinuity. Heatmaps of these two quantities over the $Q_{20}$-$Q_{30}$ plane are plotted in Fig. \ref{fig:218Ra_over_dd}, with darker areas corresponding to lower overlaps or higher density distances. When compared to heatmaps showing the values of $Q_{40}$ and $E_\mathrm{HFB}$ across the same area in Fig. \ref{fig:218Ra_q40_ener}, the results of both metrics clearly indicate a discontinuity along the diagonal ridge defined by the sharp changes in the nuclear configuration and potential energy.

\subsubsection{Parameters}

The Frontier DPM scales exponentially with the size of the search space, although far less so than a brute-force tree search. In order to reduce memory and time requirements, the 3D PES for the region of interest marked in Fig. \ref{fig:218Ra_pes} was calculated in four sub-regions, chosen to fit the discontinuity more tightly. This means that fewer points far from the discontinuity were included in the smoothing process, and that a tighter range of $Q_{40}$ values were calculated for each point, both of which reduce the complexity of the optimisation problem. \\

The four sub-regions of the 3D PES were as follows.
\begin{itemize}
\item Region 1: $Q_{20} \in [74,80] \unitsn{b}$, $Q_{30} \in [24,30] \unitsn{b^{3/2}}$, $Q_{40} \in [34,56] \unitsn{b^2}$
\item Region 2: $Q_{20} \in [80,90] \unitsn{b}$, $Q_{30} \in [26,40] \unitsn{b^{3/2}}$, $Q_{40} \in [38,74] \unitsn{b^2}$
\item Region 3: $Q_{20} \in [90,100] \unitsn{b}$, $Q_{30} \in [34,46] \unitsn{b^{3/2}}$, $Q_{40} \in [50,88] \unitsn{b^2}$
\item Region 4: $Q_{20} \in [100,110] \unitsn{b}$, $Q_{30} \in [40,52] \unitsn{b^{3/2}}$, $Q_{40} \in [62,98] \unitsn{b^2}$
\end{itemize}

The ranges of $Q_{40}$ values allowed for each sub-region were determined by taking the minimum and maximum values found in the 2D PES, and extending the range by approximately $4 \unitsn{b^2}$ in each direction. This process assumes that the smoothed surface will not vary wildly from the original in the $Q_{40}$ direction. Limiting the search space in this way reduces the size of the problem without compromising the validity of the smoothing process, as long as the selected ranges are large enough that the smoothed surface does not reach their extremal values. \\

When obtaining boundary conditions for the combined regions from the 2D PES, it was found that the maximum gradient of one step in $Q_{40}$ ($2 \unitsn{b^2}$) could not be used, as it would be inconsistent with the boundary conditions. Therefore the maximum gradient $\Delta_\mathrm{max}$ was increased to two steps ($4 \unitsn{b^2}$) for this surface, raising the branching factor (the maximum number of choices at each step) from 3 to 5. The overlap threshold was set to $\mathcal{N}_\mathrm{min} = 0.15$.

\subsubsection{Results}

\begin{figure}
\includegraphics[width=8.6cm]{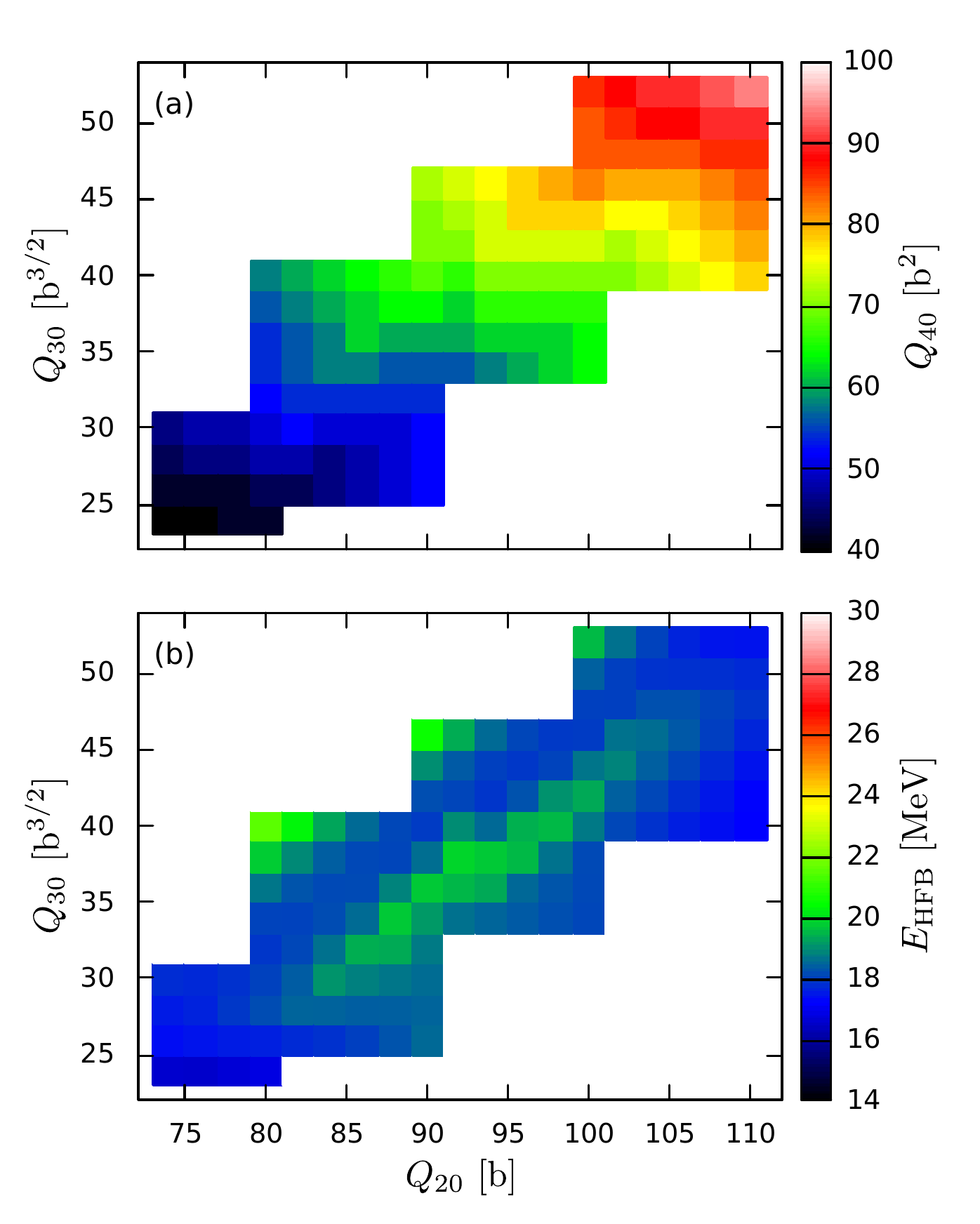}
\caption{\label{fig:218Ra_q40_ener_smoothed}Plots of the (a) HFB potential energy and (b) $Q_{40}$ as functions of $Q_{20}$ and $Q_{30}$ for the smoothed surface generated with Frontier DPM. The diagonal ridges in $Q_{40}$ and $E_\mathrm{HFB}$ visible in Figs. \ref{fig:218Ra_q40_ener}(a) and \ref{fig:218Ra_q40_ener}(b) have been removed by the Frontier DPM smoothing algorithm.}
\end{figure}

\begin{figure}
\includegraphics[width=8.6cm]{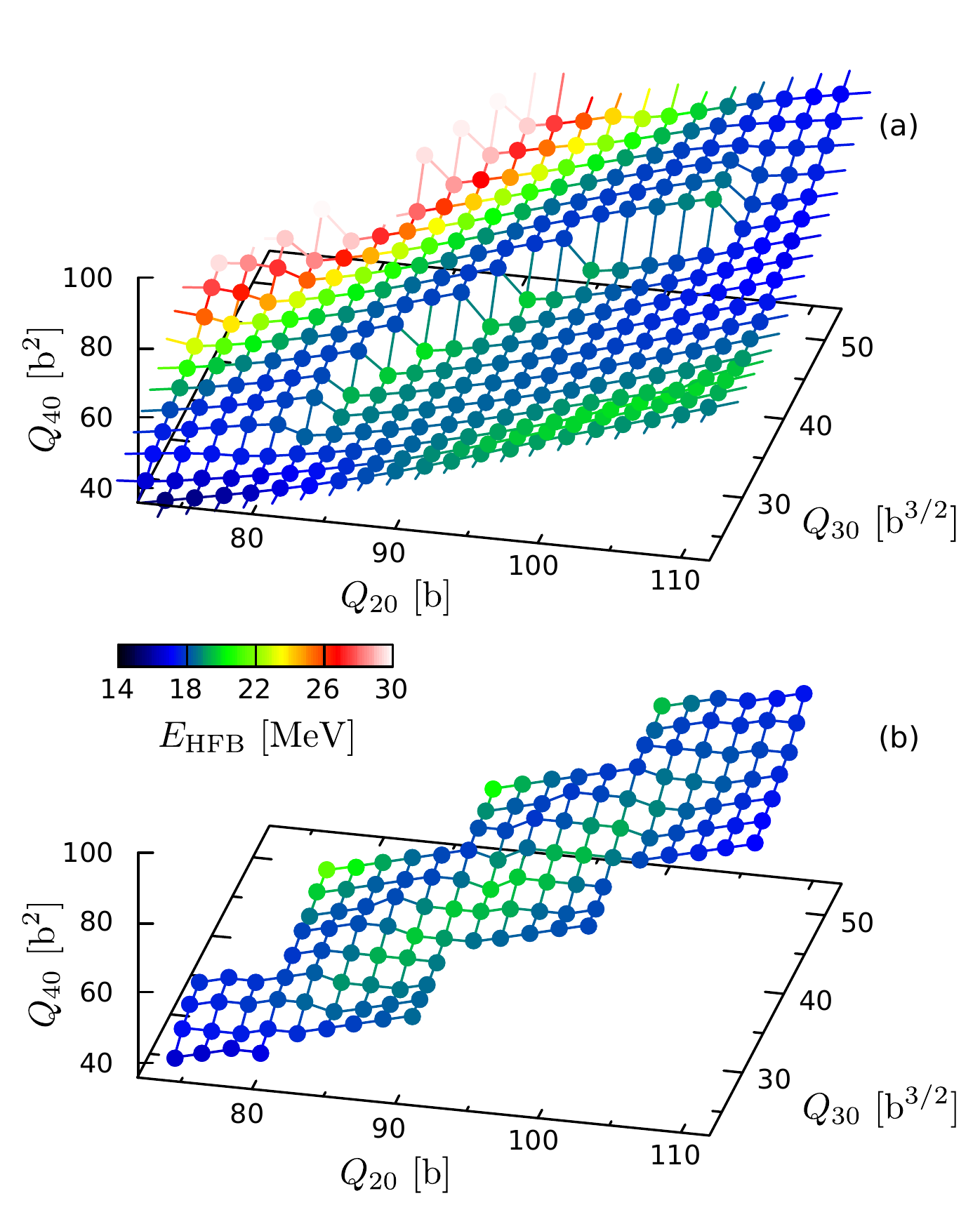}
\caption{\label{fig:218Ra_surface_3D}3D representations of the PES for $\elem{218}{}{Ra}$, in the region highlighted in Fig. \ref{fig:218Ra_pes}, showing (a) the initially calculated surface and (b) the smoothed area calculated with Frontier DPM. These display the same data as Figures \ref{fig:218Ra_q40_ener} and \ref{fig:218Ra_q40_ener_smoothed} in a way that compactly illustrates the correlations between $Q_{40}$ and $E_\mathrm{HFB}$ around the discontinuity.}
\end{figure}

Heatmaps of the $Q_{40}$ and $E_\mathrm{HFB}$ relative to the ground state in the smoothed region are shown in Fig. \ref{fig:218Ra_q40_ener_smoothed}. These demonstrate that the smoothing process has successfully eliminated the sharp discontinuity in $Q_{40}$ present in Fig. \ref{fig:218Ra_q40_ener}. The ridge in $Q_{40}$ has been removed completely, leaving the increasing trend in the positive $Q_{20}$ and $Q_{30}$ directions with only minor distortions. The information contained in these heatmaps is represented in 3D in Fig. \ref{fig:218Ra_surface_3D} to better illustrate the effects of the smoothing process. There is a region of elevated energy in the area of the discontinuity, but the ridge has been smoothed out into a gradual transition between higher-dimensional fission valleys. \\

\begin{figure}
\includegraphics[width=8.6cm]{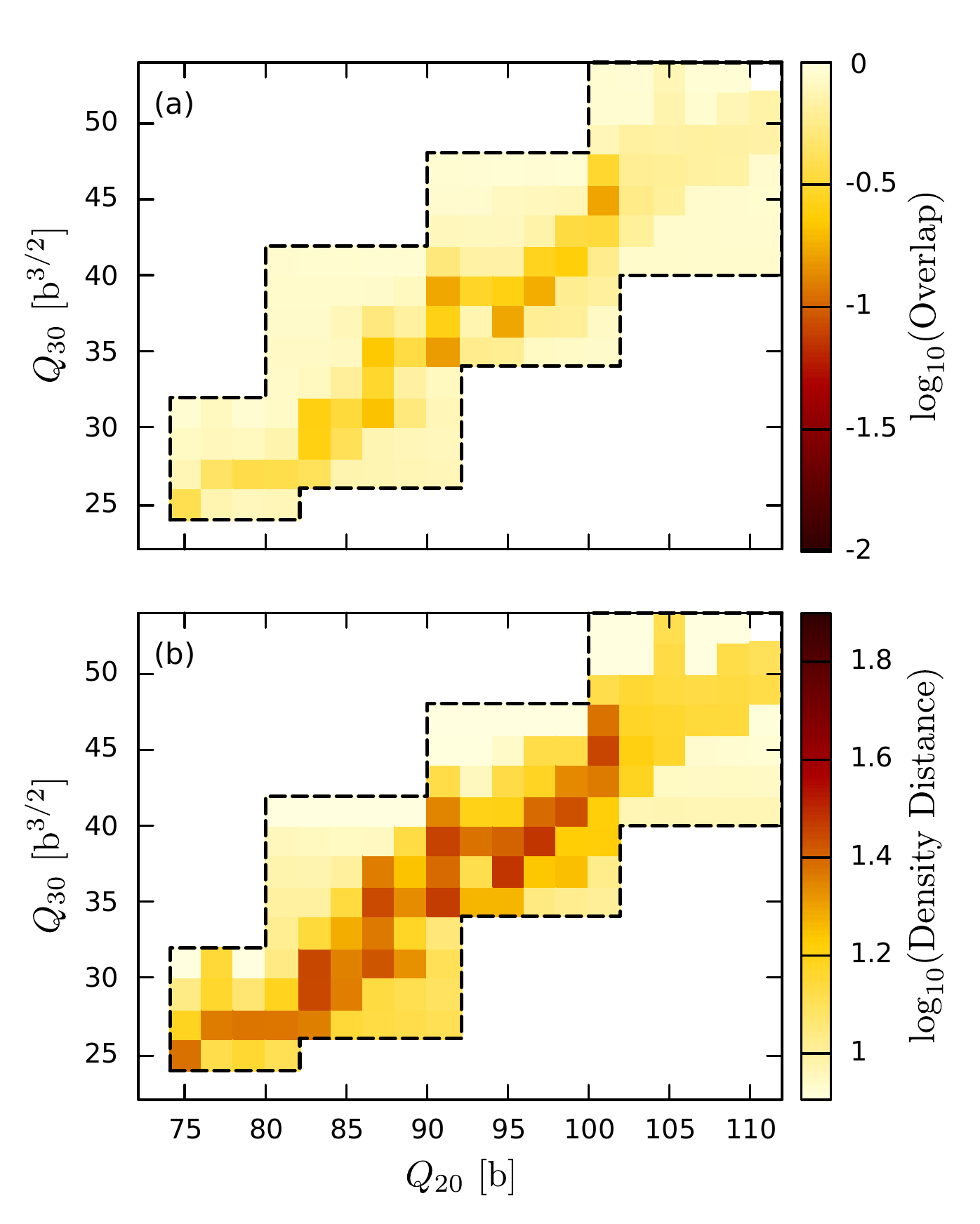}
\caption{\label{fig:218Ra_over_dd_smoothed}Logarithmic plots of the (a) overlap and (b) density distance between neighbouring points as functions of $Q_{20}$ and $Q_{30}$ for the smoothed surface. A black dotted line has been drawn to indicate the boundaries of the smoothed surface.}
\end{figure}

The overlaps and density distances for the smoothed surface shown in Fig. \ref{fig:218Ra_over_dd_smoothed} show significant improvements over the original surface, supporting the conclusion that the smoothing process has successfully removed the discontinuity from the PES. The process has achieved the desired result of eliminating zero overlaps from the PES, which makes future TDGCM calculations on this nuclide without the GOA a possibility. \\

However, it should be noted that some points of lower overlap and high density distance are close to the boundaries of the smoothed surface. It is important to be sure that the boundary conditions taken from the 2D PES are not discontinuous themselves, as this will inhibit smoothing and possibly lead to an inconsistent surface with no allowed values of $Q_{40}$ at some points. Caution should be taken when using the Frontier DPM to ensure that the boundaries of the smoothed surface are a sufficient distance from any discontinuities.

\section{\label{sec:conclusions}Conclusions}

The successful modelling of nuclear fission depends on high-quality PESs to describe the dynamics of nuclei as they proceed from their ground states to scission configurations. Several studies \cite{dubray2012,regnier2016,regnier2019,zdeb2021} have shown that discontinuities may frequently occur in PESs, obscuring the physical dynamics, preventing further analysis via time evolution, and limiting the reliability of any conclusions that can be drawn. A PES can be effectively smoothed by incorporating the dynamics of the discontinuous degrees of freedom, either by calculating a new PES in additional dimensions or by generating a higher-dimensional path over the discontinuity that can be projected onto the original PES. \\

The current work has explored the practical possibilities of PES smoothing with search-based algorithms. Modifications to the Dynamic Programming Method of Refs. \cite{baran1981,sadhukhan2013} were presented which allow it to be used to calculate the lowest-energy 1D path across a discontinuity. This search algorithm is deterministic, complete, and far more efficient than a brute-force tree search with only a few adjustable parameters. A generalised method named the Frontier DPM has also been developed, suitable for smoothing discontinuities in a 2D PES. The results from initial applications of these methods to candidate discontinuities were presented and analysed. \\

Applying the modified DPM algorithm to the discontinuity in $Q_{30}$ around the symmetric-to-asymmetric transition in the 1D fission path of $\elem{252}{}{Cf}$ has produced considerable improvements, with both overlap and density distance calculations indicating that the discontinuity was successfully eliminated. The smoothed path initially follows the adiabatic path derived from the 2D PES, producing a slightly increased barrier height as a result of maintaining continuity in the $Q_{30}$ direction. This difference highlights the importance of resolving discontinuities around fission barriers, particularly when extracting observables which are sensitive to the barrier height such as the spontaneous fission lifetime. \\

Following these results, a more thorough comparison was made between conventional adiabatic paths, linearly interpolated paths, and DPM paths, generated across the 1D discontinuity identified in $\elem{252}{}{Cf}$ as well as a second discontinuity in the 1D fission path of $\elem{222}{}{Th}$. These calculations showed that the linear interpolation method is able to produce comparable results to DPM in simple cases, but performs more poorly across discontinuities with complex shapes. Both methods generally produce paths with higher fission barriers compared to the adiabatic path. However, the linear interpolation method requires human input to choose sensible endpoints, and does not produce minimum energy solutions as should be expected under the adiabatic approximation. The version of the Dynamic Programming Method presented in this work can be considered a significant improvement over the linear interpolation method as it produces deterministic, repeatable results (given a small set of parameters) derived from physically motivated constraints on both smoothness and adiabaticity. \\

An initial survey of the $\elem{218}{}{Ra}$ PES with $N_\perp = 14,\ N_z = 21$ produced misleading results containing ``fake'' discontinuities arising from large variations in the length parameters of the harmonic oscillator basis. As a result, the PES was recalculated using $N_\perp = 15,\ N_z = 22$ and fixed oscillator lengths, revealing a genuine discontinuity in the $Q_{40}$ direction suitable as a candidate for smoothing. Applying Frontier DPM successfully removed the discontinuity, significantly increasing overlaps and decreasing density distances in the vicinity. These results show that smoothing techniques will be a helpful tool for the authors' future goal of performing TDGCM calculations without the GOA, as discontinuous regions with zero overlaps which would block time evolution can be handled in a process that is physically and computationally reasonable. \\

The algorithms presented in this work are simple to implement, and have a reasonable computational cost for discontinuities in a single coordinate occurring in 1D paths or 2D surfaces. However, it would be na\"ive to assume that every discontinuity can be resolved with two or three-dimensional calculations. To smooth more complex discontinuities in higher-order multipole moments, it is likely that search-based approaches will have to be abandoned in favour of methods which can leverage prior knowledge of the PES. In particular, DPM and Frontier DPM do not take advantage of information about the local minima corresponding to different fission channels available to the nucleus. Although the HFB method's strict adherence to these valleys is the primary cause of discontinuities in the first place, it may be fruitful to consider an iterative approach that can gradually smooth out the initial PES, rather than blindly searching the entire coordinate space. The continued development of physically justified, efficient methods to model the energy and deformation of nuclei, whether in this direction or others, will undoubtedly yield new and deeper insights into the dynamics of nuclear fission.

\begin{acknowledgments}
The authors thank P. McGlynn and L. M. Robledo for helpful discussions. \\

This work has been supported by the Australian Research Council under Grant No. DP190100256. Calculations were performed using computational resources provided by the Australian Government through the National Computational Infrastructure (NCI) under the ANU Merit Allocation Scheme. \\

N.-W.T.L. acknowledges the support of the Australian Commonwealth through the Australian Government Research Training Program (AGRTP) Scholarship.
\end{acknowledgments}

\bibliography{Smoothing_Paper_2021_NWTLau}

\end{document}